\documentclass[aps,pre,superscriptaddress,balancelastpage,showpacs,reprint,floatfix]{revtex4-2}

\usepackage{blindtext}
\usepackage{centernot}
\usepackage{graphicx}
\usepackage{amsmath,bbold}
\usepackage{times}
\usepackage{amssymb}
\usepackage{mathrsfs}
\usepackage{amsthm}
\usepackage{chemarr}
\usepackage{xcolor}
\usepackage{url}
\usepackage{version}
\usepackage{bbm}

\definecolor{darkgreen}{RGB}{0,100,0}

\usepackage{tikz}
\usetikzlibrary{decorations.pathmorphing}
\usetikzlibrary{arrows.meta}
\usetikzlibrary{calc}
\usepackage[percent]{overpic}
\usepackage{bm}
\definecolor{linkcolor}{rgb}{0,0,0.6}

\newcommand{\bp}{{\bf p}}
\newcommand{\q}{{\bf q}}
\newcommand{\br}{{\bf r}}
\newcommand{\bv}{{\bf v}}

\newcommand{\x}{{\mathbf{x}}}
\newcommand{\cL}{\mathcal{L}}
\newcommand{\cO}{\mathcal{O}}
\newcommand{\cH}{\mathcal{H}}
\newcommand{\cP}{\mathcal{P}}

\newcommand{\bu}{\mathbf{u}}
\newcommand{\mR}{\mathbb{R}}
\newcommand{\blambda}{\mathbf{\Lambda}}

\newcommand{\eps}{\varepsilon}

\newcommand{\G}{\text{\tiny G}}

\newcommand{\eq}{\mathrm{eq}}
\newcommand{\bare}{^{(0)}}

\newcommand\hnp[2]{h^{\{#1\}}_{#2}}
\newcommand\anp[2]{a^{\{#1\}}_{#2}}

\usepackage{lipsum}
\usetikzlibrary{patterns}

\usepackage[pdftex]{hyperref}
\graphicspath{{../}}
\begin{document}

\title{Statistical mechanics of a cold tracer in a hot bath}

\author{Amer Al-Hiyasat}
\affiliation{Department of Physics, Massachusetts Institute of Technology, Cambridge, Massachusetts 02139, USA}
\author{Sunghan Ro}
\affiliation{Department of Physics, Harvard University, Cambridge, Massachusetts 02138, USA}

\author{Julien Tailleur}
\affiliation{Department of Physics, Massachusetts Institute of Technology, Cambridge, Massachusetts 02139, USA}

\date{\today}
\begin{abstract}
We study the dynamics of a zero-temperature particle interacting linearly with a bath of hot Brownian particles. Starting with the most general model of a linearly-coupled bath, we eliminate the bath degrees of freedom exactly to map the tracer dynamics onto a generalized Langevin equation, allowing for an arbitrary external potential on the tracer. We apply this result to determine the fate of a tracer connected by springs to $N$ identical bath particles or inserted within a harmonic chain of hot particles. In the former ``fully-connected" case, we find the tracer to transition between an effective equilibrium regime at large $N$ and an FDT-violating regime at finite $N$, while in the latter ``loop" model the tracer never satisfies an FDT. We then study the fully-connected model perturbatively for large but finite $N$, demonstrating signatures of irreversibility such as ratchet currents, non-Boltzmann statistics, and positive entropy production. Finally, we specialize to harmonic external potentials on the tracer, allowing us to exactly solve the dynamics of both the tracer and the bath for an arbitrary linear model. We apply our findings to show that a cold tracer in a hot lattice suppresses the fluctuations of the lattice in a long-ranged manner, and we generalize this result to linear elastic field theories.
\end{abstract}
\maketitle

\tableofcontents{}
\section{Introduction} 
\label{sec:introduction}

Studying the random motion of tracer particles in fluids has been a long-standing topic of interest since the early work of Brown and Einstein.
Much progress has been made by studying simple models in which the tracer dynamics can be coarse-grained exactly into a generalized Langevin equation. For Hamiltonian systems, this approach was pioneered by Feynman and Vernon~\cite{feynman1963}, Ford, Kac and Mazur~\cite{fkm1965}, as well as Caldeira and Leggett~\cite{caldeira1983}, and generally involves modelling the  bath as a collection of harmonic oscillators whose dynamics can be integrated out exactly.

In recent years, a growing body of work has focused on the behavior of tracers within nonequilibrium baths~\cite{wu2000,Loi2008,Underhill2008,Leptos2009,Dunkel2010,Kurtuldu2011,Mino2011,Zaid2011,wilson2011differential,Argun2016,Foffano2012,Mino2013,Kasyap2014,Morozov2014,Thiffeault2015,Suma2016,Burkholder2017, Pietzonka2018,cui2018generalized, Chaki2018,Kanazawa2020,Knezevic2020,maes2020,Abbaspour2021,Reichert2021, granek2022, solon2022, santra2023dynamical, pelargonio2023generalized,sarkar2024harmonic}. Of particular interest is the case of multi-temperature mixtures, which have attracted attention due to their rich single- and many-body phenomenology~\cite{grosbergjoanny2015, weberfrey2016, tanaka2017, ilkerjoanny2021,grosberg2021, jardat2022diffusion, goswami2023trapped, burov2024, Damman2024, santra2024forces}. 
In a companion Letter~\cite{letter}, we studied the behavior of a single cold particle within a bath of hot Brownian particles, 
revealing a rich dynamics for the cold tracer that straddles equilibrium and nonequilibrium regimes. We found that the tracer equilibrates at the temperature of the hot bath only in the limit where it interacts with infinitely many bath particles. 
At finite bath densities, the tracer falls out of equilibrium and displays ratchet currents, boundary accumulation, and a positive entropy production rate. These results were obtained from numerical simulations of particles interacting via short-ranged pairwise forces. 
To account for our results analytically, we showed that they are well reproduced by linear bath models which are amenable to theoretical treatment. Given the prominent role played by such linear models in the context of Hamiltonian systems, we believe that they are worthy of study in their full generality, which is the aim of this article.

We thus consider systems in which a zero-temperature, overdamped tracer at position $x$ interacts with a bath of Brownian colloids $(x_1, \dots, x_N)$ at temperature $T>~0$:
\begin{subequations}
\label{eq:genEOM}
        \begin{eqnarray}
    \dot{x} &=& -\mu U'(x) -\mu \partial_x V(x,\{x_i\}),\\
    \dot{x}_i &=& -\mu_i \partial_i V(x,\{x_i\}) + \sqrt{2\mu_i T} \eta_i(t).
\end{eqnarray}
\end{subequations}
In the above, $V$ is a quadratic interaction potential, $U$ is an external potential on the tracer, the $\{\mu_i \}$ are the particle mobilities, and the $\{\eta_i \}$ are independent centered unit Gaussian white noises. Spatial derivatives are indicated with primes for univariate functions and partial derivatives for multivariate ones. 
In section~\ref{sec:genlan}, we derive exactly the generalized Langevin equation of the tracer for an arbitrary linearly-coupled bath. 
We apply this result to two specific examples; first, a fully-connected model (Fig.~\ref{fig:examples}a) where the cold tracer is connected by springs to $N$ bath particles, which can be thought of as a mean-field model of a fluid bath. Second, we consider a model in which the cold tracer is inserted within a loop of $N$ hot particles connected by springs (Fig.~\ref{fig:examples}b), mimicking a gel or a solid. We find the fully-connected model to transition between active and effective equilibrium dynamics depending on $N$, while the loop model always remains out of equilibrium. We then, in section~\ref{sec:active}, develop a perturbation theory to study the tracer dynamics in the fully connected model for large but finite $N$, characterizing its departure from equilibrium and the emergence of irreversibility. 
To characterize the dynamics of both the tracer and the bath, we turn in section~\ref{sec:exact} to the exactly solvable case where the external potential $U$ is also quadratic. This allows the distributions of both the tracer and the bath particles to be computed exactly for finite $N$. We find that, in the loop model, the cold tracer has a long-ranged effect on the bath. Finally, we generalize this result to higher-dimensional lattices and to more general elastic models in section~\ref{sec:fieldtheory} using a field-theoretic approach. 

\begin{figure} 
\begin{tikzpicture}[scale=1]
\def\h{1.73}
\fill[rounded corners=5pt, blue ,opacity=0.2] (-0.75,-0.6+\h) rectangle (0.75,0.75+\h);
\fill[rounded corners=5pt, brown,opacity=0.2] (-1.57,-.47) rectangle (1.57,0.4);
\node[brown, font=\footnotesize] at (0,-0.7) {$T>0$};
\node[blue, font=\footnotesize] at (0,0.9+\h) {$T=0$};
\draw[scale=0.25,domain=-2.23:2.36,smooth,variable=\x,red, shift={(0,\h+4.15)}, line width = 0.7pt] plot ({\x}, {\x*\x*\x*\x/4 - \x*\x/1.5 - \x/4});

\node[red, font=\footnotesize] at (0,0.42+\h) {$U(x)$};

\foreach \i/\x in {1/-1.2, 2/-0.45, 3/0.3, {\scriptscriptstyle \! N}/1.2} {
    \draw[decorate,decoration={coil,aspect=0.5,amplitude=1.3pt,pre length=0.35cm,post length=0.5cm,segment length=1mm}] (\x,0) -- (0,\h);
    \path (\x-0.55,0) -- (0.1,\h-0.5);
    \draw[fill=white] (\x,0) circle [radius=0.23] node[font=\footnotesize] {$x_{\i}$} ;
}
  \node[font=\footnotesize] at (0.77,0) {$\dots$};
    \node[font=\footnotesize] at (0.87,0.77) {$k$};
    \node[font=\footnotesize] at (0.32,0.77) {$k$};
    \node[font=\footnotesize] at (-0.42,0.77) {$k$};
    \node[font=\footnotesize] at (-0.89,0.77) {$k$};
\draw[fill=white] (0,\h) circle [radius=0.23] node[font=\footnotesize] {$x$};



 \node at (-1.42, 0.86+\h) {(a)};
\end{tikzpicture} 
\hspace{2em}
\raisebox{1.24em}{
    \begin{tikzpicture}[scale=1]
\def\h{1.7}

\fill[brown,opacity=0.2] 
    plot [smooth cycle] coordinates {(.7,-0.5) (1.05,-0.3) (1.35,0.1) (1.4,0.9)  (1.1,1.60) (.7,1.58) (.4,1.07) (-.4,1.07) (-.7,1.58) (-1.1,1.60) (-1.4,0.9) (-1.35,0.1)  (-1.05,-0.3) (-.7,-0.5) (0,-0.6)};

\fill[blue,opacity=0.2] 
    plot [smooth cycle] coordinates {(0.7, 2.3)  (-0.7, 2.3) (-0.4, 1.2) (0.4, 1.2)};
    
\node[brown, font=\footnotesize] at (0,0.5) {$T>0$};
\node[blue, font=\footnotesize] at (0,0.85+\h) {$T=0$};
\draw[scale=0.25,domain=-2.23:2.36,smooth,variable=\x,red, shift={(-0.2,\h+3.9)}, line width = 0.7pt] plot ({\x}, {\x*\x*\x*\x/4 - \x*\x/1.5 - \x/4});

\node[red, font=\footnotesize] at (0,0.45+\h) {$U(x)$};

\def\xOne{0.9}
\def\yOne{1.3}
\def\xTwo{1.1}
\def\yTwo{0.6}
\def\xThree{0.75}
\def\yThree{-0.1}
\def\yZero{\h}

\foreach \xstart/\ystart/\xend/\yend in {0/\h/\xOne/\yOne, \xOne/\yOne/\xTwo/\yTwo, \xTwo/\yTwo/\xThree/\yThree, 0/\h/-\xOne/\yOne, -\xOne/\yOne/-\xTwo/\yTwo, -\xTwo/\yTwo/-\xThree/\yThree} {
    \draw[decorate,decoration={coil,aspect=0.5,amplitude=1.1pt,pre length=0.22cm,post length=0.15cm,segment length=0.9mm}] (\xstart,\ystart) -- (\xend,\yend);
}

\draw[decorate, decoration={coil,aspect=0.5,amplitude=1.3pt,pre length=0.25cm,post length=.02cm,segment length=0.9mm}] (\xThree,\yThree) -- (.32,\yThree-0.3);
\draw[decorate, decoration={coil,aspect=0.5,amplitude=1.3pt,pre length=0.25cm,post length=.02cm,segment length=0.9mm}] (-\xThree,\yThree) -- (-0.32,\yThree-0.3);

\node[font=\small] at (0, \yThree-0.35) {$\dots$};

\foreach \i/\x/\y in {1/\xOne/\yOne, 2/\xTwo/\yTwo, 3/\xThree/\yThree, {\scriptscriptstyle \!N}/-\xOne/\yOne, \scalebox{0.78}{$\scriptscriptstyle \mkern-4.5mu N\mkern-1.3mu \text{--}\mkern-1.2mu 1$}/-\xTwo/\yTwo, \scalebox{0.78}{$\scriptscriptstyle \mkern-4.5mu N\mkern-1.3mu \text{--}\mkern-1.2mu 2$}/-\xThree/\yThree} {
    \draw[fill=white] (\x,\y) circle [radius=0.25] node[font=\footnotesize] {$x_{\i}$} ;
}
\draw[fill=white] (0,\yZero) circle [radius=0.25] node[font=\footnotesize] {$x$};
\node at (-1.3,\h+0.82) {(b)};
\end{tikzpicture}}

\caption{A zero-temperature tracer in an external potential $U(x)$ coupled to two models of a linearly-interacting hot bath. (a) Fully-connected model: The tracer is connected by linear springs to all bath particles. (b)  Loop model: The tracer is inserted within a harmonic chain of hot particles. These are two instances of the more general coupling described by Eq.~\eqref{eq:genlanham} and studied throughout the article.
}\label{fig:examples}
\end{figure}

\section{Generalized Langevin equation of a cold tracer in a linearly-coupled bath} \label{sec:genlan}
In this section, we show how the bath degrees of freedom can be  eliminated from Eqs.~\eqref{eq:genEOM} to obtain a generalized Langevin equation for the tracer, so long as all interactions among particles, $V(x, \{x_i\})$, are quadratic. We first derive general expressions for the retarded friction and noise correlations of a tracer in an arbitrary linearly-coupled bath. We then compute these explicitly for the fully connected model shown in Fig.~\ref{fig:examples}(a) (Sec.~\ref{subsec:genLanMF}), and for the loop model shown in Fig.~\ref{fig:examples}(b) (Sec.~\ref{sec:loopmodel}). 
We allow for an arbitrary external potential $\tilde{U}(x)$, not necessarily quadratic, acting on the tracer. We work in one spatial dimension for simplicity; Appendix~\ref{appen:ddimgenlan} provides the generalization to $d$-dimensions. It also covers tracers with a positive temperature $T_0$.

Letting $\x \equiv (x_1, \dots, x_N)$ denote the vector of bath variables and $x$ the tracer position, we consider the potential energy
\begin{equation}\label{eq:genlanham}
\mathcal{H}(x, \mathbf{x}) = \frac{1}{2} \mathbf{x}^T A \mathbf{x} + x \mathbf{c} \cdot \mathbf{x} + \tilde{U}(x),\end{equation}
where $A$ is a symmetric positive-definite matrix and $\mathbf{c} \in \mathbb{R}^N$ is a vector of coupling coefficients. For ease of notation, we restrict the bath particle mobilities to be equal, $\mu_1 = \dots = \mu_N$, and choose time units such that $\mu_{i} = 1$. The generalization to arbitrary mobilities is straightforward. The evolution equations then read:
\begin{align}
    \dot{x} &= -\mu \tilde{U}'(x) - \mu\mathbf{c}\cdot \mathbf{x}, \label{eq:tracerGeneralLinear}
    \\
    \dot{\mathbf{x}} &= -A \mathbf{x} - x\mathbf{c}+\sqrt{2T} \bm{\eta}(t),
\end{align}
where $\langle \eta_i(t) \eta_j(t') \rangle = \delta_{ij}\delta(t-t')$. The second of these equations can be integrated to obtain the bath dynamics as a function of the tracer trajectory
\begin{equation}\label{eq:formalsol}
\mathbf{x}(t) = e^{-At}\mathbf{x}(t_0) + \int_{t_0}^t ds e^{-A(t-s)}\left[\sqrt{2T} \bm{\eta} (s) - \mathbf{c} x(s) \right].
\end{equation}
Taking $t_0=-\infty$ and substituting Eq.~\eqref{eq:formalsol} into Eq.~\eqref{eq:tracerGeneralLinear} leads to a closed evolution equation for the tracer:
\begin{equation}\label{eq:qdot}
\dot{x} = -\mu \tilde{U}'(x) + \mu \int_{-\infty}^t ds \mathbf{c}\cdot e^{-A(t-s)} \mathbf{c} x(s) + \mu \xi(t),\end{equation}
where $\xi$ is an Ornstein-Uhlenbeck noise with correlations,
\begin{equation}
    G(s) \equiv \langle \xi(t+s)\xi(t) \rangle = T \mathbf{c}\cdot A^{-1}e^{-A|s|}\mathbf{c}. \label{eq:noisekern}
\end{equation}
To turn Eq.~\eqref{eq:qdot} into a generalized Langevin equation, we write $e^{-A(t-s)}$ as $A^{-1} \partial_s e^{-A(t-s)}$ and integrate by parts~\footnote{The boundary term at $s=t$ leads to the second term in the right-hand side of Eq.~\eqref{eq:IBPs} while the boundary at $s=-\infty$ vanishes since $A$ is positive definite.}:
\begin{align}
\dot{x} = -& \mu \tilde{U}'(x) + \mu\mathbf{c} \cdot A^{-1} \mathbf{c} x(t) \nonumber\\&-\mu \int_{-\infty}^t ds \mathbf{c}\cdot A^{-1}e^{-A(t-s)}\mathbf{c}\dot{x}(s) + \mu \xi(t)\label{eq:IBPs}
\end{align}
Replacing $x(t)$ with $\int^t ds \dot{x}(s)$, we arrive at
\begin{equation}\int_{-\infty}^t ds K(t-s) \dot{x}(s) = -\tilde{U}'(x(t))+\xi(t), \label{eq:genlan}\end{equation}
where
\begin{equation}K(s) = 2\mu^{-1}\delta(s) + \mathbf{c} \cdot A^{-1}(e^{-A|s|}-1)\mathbf{c}. \label{eq:frickern}
\end{equation}
The tracer is then endowed with a \textit{bona fide} equilibrium dynamics whenever the friction $K(s)$ and the noise kernel $G(s)$ satisfy a fluctuation-dissipation theorem (FDT): $G(s) = T K(s)$ \cite{zwanzig2001}. In this case, the tracer becomes Boltzmann-distributed in the steady state, with $p(x) \propto \exp(-\tilde{U}(x)/T)$. We next determine the conditions for an FDT to be satisfied in the fully-connected and loop models of Fig.~\ref{fig:examples}.

\subsection{Fully-connected model} \label{subsec:genLanMF}
We consider the fully-connected model of Fig.~\ref{fig:examples}(a), which can be interpreted as a mean-field model of a tracer immersed in a cold fluid containing colloidal particles at an enhanced temperature. We first apply the formalism introduced above to construct the generalized Langevin equation describing the dynamics of this system (see Ref.~\cite{grosberg2021} for related results in the context of multi-temperature systems).

For simplicity, we consider a noninteracting bath, though our results are general and apply in the presence of an interaction potential between the bath particles, as shown in Sec.~\ref{sec:active}. For this model, the potential energy takes the form
\begin{equation}\cH = \sum_{i=1}^N \frac{k}{2}(x_i-x)^2 + U(x)\;.\label{eq:hamfullycon}
\end{equation}
Matching Eq.~\eqref{eq:hamfullycon} to Eq.~\eqref{eq:genlanham} leads to $A_{ij} = k\delta_{ij}$, $c_i = -k$, and $\tilde{U}(x) = U(x) + Nkx^2/2$. Since we would like a generalized Langevin equation involving $U$ and not $\tilde U$, we note that Eq.~\eqref{eq:genlan} is invariant under the combined transformation $\tilde{U}(x) \rightarrow \tilde{U}(x) + bx^2/2$ and $K(s) \rightarrow K(s) - b$. Choosing $b=-Nk$, we can thus eliminate the term $N k x^2/2$ in $\tilde U$ to obtain a generalized Langevin equation like Eq.~\eqref{eq:genlan}, with $\tilde U=U$ and
\begin{subequations}
 \label{eq:unrescaledkernels}
\begin{align}
    K(s) &= 2\mu^{-1}\delta(s) + Nke^{-k|s|}, \label{eq:connectedFricKern}\\
    G(s) &= T N k e^{-k|s|}. \label{eq:connectedNoiseKern}
\end{align}
\end{subequations}
 Generically, the first term of Eq.~\eqref{eq:connectedFricKern} breaks the FDT, $K(s) \neq T G(s)$, so that the tracer follows a nonequilibrium dynamics. We show in Sec.~\ref{sec:active} that this FDT violation leads to a persistent motion reminiscent of an active particle, and that the tracer exhibits boundary accumulation, ratchet currents, and entropy production.

There are, however, some limiting cases of interest in which an FDT is recovered. For example, if the springs are very stiff, $k \rightarrow \infty$, we may use $\lim_{a\rightarrow \infty} a e^{-a |s|} = 2\delta(s)$ to show that the kernels~\eqref{eq:unrescaledkernels} become white:
\begin{subequations}\label{eq:largeckernels} \begin{align}K(s) &\rightarrow 2(\mu^{-1}+N) \delta(s), \\ G(s) &\rightarrow 2 T N\delta(s).\end{align}
\end{subequations}
This implies an equilibrium dynamics at an effective temperature $T_\mathrm{eff} = T \mu N/(1+\mu N)$. In this limit, tracer and bath particles behave as a single rigid body, with a drag coefficient $\mu^{-1} + N$ and a temperature lowered from $T$ by the cold tracer.

It is of experimental interest, particularly in the study of colloids in active enzyme solutions \cite{ghosh2021, zhao2017enhanced}, to study the behavior of a tracer that interacts with very many bath particles at once.
To characterize this large-$N$ limit, we first note that a rescaling of time $t\rightarrow t/g$ in Eq.~\eqref{eq:genlan} is equivalent to the replacement $K(s) \rightarrow K(g s), G(s) \rightarrow G(g s)$. Making the choice $g = N$, the rescaled kernels are
\begin{subequations}
\begin{align}
    K(s) &= \frac{2}{\mu N}\delta(s) + N k e^{-Nk |s|}, \\
    G(s) &= T N k e^{-N k |s|}.
\end{align}
\end{subequations}
If $N$ is made large while keeping the other parameters $\mathcal{O}(N^0)$, the rescaled kernels become white 
\begin{equation}K(s) \rightarrow 2 \delta(s), \quad G(s) \rightarrow 2 T \delta(s), \quad (N\gg 1). \label{eq:largeNkernels}\end{equation} 
We thus have an equilibrium dynamics at temperature $T$ for large $N$. Although there continues to be a heat flux from the $T>0$ to the $T=0$ reservoir, the influence of the cold reservoir vanishes in the large $N$ limit, and the tracer equilibrates at the temperature of the hot bath. 

In experiments, increasing the number of hot particles that interact with the tracer can be achieved by increasing either the particle density or the tracer radius. In a realistic fluid, the latter also reduces the tracer mobility, which calls for a study of the joint limit $N \rightarrow \infty, \mu \rightarrow 0$.  The result here depends on how quickly one sends $\mu$ to zero; letting $\mu$ take the large-$N$ form $\mu\sim \tilde{\mu} N^{-\alpha}$ with $\tilde{\mu} \in \mathcal{O}(1)$ and $\alpha>0$, we find that if $\alpha>1$, the tracer no longer equilibrates: it ceases to fluctuate and follows a zero-temperature gradient descent in $U(x)$. If $\alpha<1$, the tracer equilibrates at temperature $T$ for large $N$. Interestingly, in the marginal case $\alpha=1$, the tracer equilibrates at a tunable temperature:
\begin{equation} 
T_{\mathrm{eff}} = \frac{\tilde{\mu}}{1+\tilde{\mu}} T.
\end{equation}
For a spherical tracer of radius $R$ interacting with hot Brownian colloids in a fluid, the Stokes formula predicts $\mu \propto R^{-1}$. 
Since the number of colloids interacting with the tracer scales with its surface area, $N \propto R^2$, we have $\alpha=1/2$ and thus expect an asymptotically large tracer to equilibrate at temperature $T$.

Another regime of interest, which has drawn attention in the context of Hamiltonian systems~\cite{feynman1963, caldeira1983}, is the so-called weak-coupling limit, in which the number of bath particles is large but the coupling to the tracer is vanishingly weak. To explore this regime, we let $k \sim \tilde{k} N^{-\alpha}$. When $\alpha < 1$, we recover Eq.~\eqref{eq:largeNkernels} upon rescaling time by $N$, implying equilibration at temperature $T$. When $\alpha > 1$, Eqs.~\eqref{eq:unrescaledkernels} show that the kernels approach $K(s) = 2\mu^{-1} \delta(s)$ and $G(s) = 0$, implying a zero-temperature overdamped dynamics for the tracer. If $\alpha = 1$, the friction and noise kernels~\eqref{eq:unrescaledkernels} do not satisfy an FDT, and the tracer dynamics remain out of equilibrium as $N\rightarrow \infty$. \if{ 
\begin{subequations}
\begin{align}
K(s) &\rightarrow \frac{2}{\mu N} \delta(s) + \tilde{k} e^{-\tilde{k}|s|}, 
\\G(s) &\rightarrow T\tilde{k} e^{-\tilde{k}|s|}.
\end{align}
\end{subequations}
If we naively set $N = \infty$ in the above, it appears that we have an FDT at temperature $T$. However, this limit is singular because the factor $N^{-1}$ multiplies a delta function. It turns out that this case has a large-$N$ limit which is out of equilibrium and resembles an active particle. We will return to this $\alpha=1$ case in section~\ref{sec:active}.}\fi

Before concluding this section, we mention here that by an appropriate scaling of parameters, one can obtain from this fully-connected model a known model of active particles. Specifically, if, for some number $\nu > 0$, the parameters are scaled as
\begin{equation}k \sim \tilde{k}N^{-1-\nu}, \quad \mu\sim\tilde{\mu}N^{-1-\nu}, \quad T \sim \tilde{T}N^\nu.\end{equation}
Then upon rescaling $t \rightarrow N^{-1-\nu} t$ and taking $N$ large, we find
\begin{subequations}
    \begin{align}
        K(s) &= 2 \tilde{\mu}^{-1} \delta(s) \\
        G(s) &= 2 \tilde{T} \tilde{k} e^{-\tilde{k} |s|}
    \end{align}
\end{subequations}
When observed on (unrescaled) time scales $t \sim N^{1+\nu}$, the tracer thus behaves as an Active Ornstein-Uhlenbeck Particle (AOUP), which is one of the standard models of self-propelled motion. Although this scaling may appear artificial, it establishes that the heat flux between the hot and cold reservoirs can be harnessed to produce an active dynamics that is known to display signatures of irreversibility. In section~\ref{sec:active}, this connection to active matter will be shown to hold generically at finite $N, c$ and $T$, although the dynamics then constitute a novel type of active particle that is different from the AOUP.

\subsection{The loop model}\label{sec:loopmodel}
We now turn to the loop model of Fig.~\ref{fig:examples}(b), in which a cold particle is inserted within a one-dimensional lattice of hot particles~\cite{fkm1965, rieder1967properties}. In this case, the number of bath particles can be made large without altering the connectivity of the tracer to the bath. This model, together with its generalization to higher dimensions (Sec.~\ref{sec:fieldtheory}), can thus be thought to describe a cold tracer embedded within a gel---or any other elastic medium---whose fluctuations are enhanced by an active drive, such as in a cytoskeletal network ~\cite{juelicher2007active,mizuno2007nonequilibrium,wilhelm2008out,fodor2014energetics,ben2015modeling,bohec2019distribution} or an active solid \cite{massana2024multiple, baconnier2022selective}. The model is defined by the Hamiltonian
\begin{equation}
\cH = U(x) + \frac{1}{2}\sum_{i=0}^{N} (x_i - x_{i+1})^2\;,
\end{equation}
where we implicitly identify $x_0 \equiv x_{N+1} \equiv x$, and where the spring stiffnesses have been set to unity without loss of generality. Comparing to Eq.~\eqref{eq:genlanham}, we find $c_i = -(\delta_{i 1} +~\delta_{iN})$, $A_{ij} = 2 \delta_{ij} - \delta_{i, j \pm 1}$, and $\tilde{U}(x) = U(x) + kx^2$. To obtain a generalized Langevin equation, we use Eqs.~\eqref{eq:noisekern} and~\eqref{eq:frickern} and absorb the term $k x^2$ into the friction kernel as before. This leads to
\begin{align}
    K(s) &= 2 \delta(u) + 2 \left[(A^{-1} e^{-A|s|})_{11} + (A^{-1} e^{-A|s|})_{1N}\right], \nonumber\\
    G(s) &= 2 T \left[(A^{-1} e^{-A|s|})_{11} + (A^{-1} e^{-A|s|})_{1N}\right],\label{eq:fkmrawkernels}
\end{align}
where, in deriving the above, we have used the symmetry $(A^{-1} e^{-Au})_{11} = (A^{-1} e^{-Au})_{NN}$. We will show that these kernels have a well-defined $N\rightarrow \infty$ limit, but that, unlike in the fully-connected model, this limit does not satisfy an FDT. 

To characterize the large-$N$ limits of $K$ and $G$ requires us to compute terms of the form $\lim_{N\rightarrow \infty}(A^{-1} e^{-Au})_{ij}$. To do this, we first diagonalize $A \equiv P \Lambda P^{T}$, where $P$ is the orthogonal matrix whose columns are the eigenvectors $\{\mathbf{v}^{(n)}\}$ of $A$. We then find
\begin{equation}
(A^{-1} e^{-Au})_{ij} = \frac{P_{ik}P_{jk} e^{-\lambda_k u}}{\lambda_k}\;,
\end{equation}
where $\{\lambda_n\}$ denotes the eigenvalues of $A$.
The eigendecomposition of a tridiagonal Toeplitz matrix like $A$ follows known formulae, leading to:
\begin{subequations}\begin{align}
\lambda_n &= 2 - 2 \cos\left(\frac{n \pi}{N+1}\right), \\
v_k^{(n)} &= \sqrt{\frac{2}{N+1}}\sin \left(\frac{n k \pi}{N+1}\right),
\end{align}\end{subequations}
which can be understood by interpreting $A$ as a discrete Laplace operator. Writing $P_{ij}=v_j^{(i)}$, we find
\begin{equation} (A^{-1} e^{-A|s|})_{ij} = \sum_{k=1}^N \frac{\sin \frac{k i \pi}{N +1} \sin \frac{k j \pi}{N+1} e^{\left(2\cos \frac{k \pi}{N+1}-2\right) |s|}}{(N+1)\left(1-\cos\frac{k \pi}{N+1}\right)} . \end{equation}
The entries which contribute to Eq.~\eqref{eq:fkmrawkernels} are $i = j =1$ and $i=1$, $j=N$. The latter vanishes in the large $N$ limit; this is because $\sin[k\pi/(N+1)] = (-1)^{k+1} \sin[k N\pi/(N+~1)]$, so that $(A^{-1} e^{-A|s|})_{1N}$ is a sum of alternating positive and negative terms that cancel as $N\rightarrow \infty$. On the other hand, $(A^{-1} e^{-Au})_{11}$ is a sum of positive terms that simplifies to
\begin{equation}(A^{-1} e^{-A|s|})_{11} = \sum_{k=1}^N \frac{e^{\left(2\cos\!\frac{k \pi}{N+1} - 2 \right) |s|}}{N+1}\left[1+\cos\frac{k \pi}{N+1}\right] . \label{eq:A11sum}\end{equation}
Introducing $y=k/(N+1)$, we identify the large-$N$ limit of Eq.~\eqref{eq:A11sum} with the integral:
\begin{equation}(A^{-1} e^{-A|s|})_{11} = e^{-2|s|} \int_0^1 dy \left[1 + \cos(\pi y)\right] e^{2 \cos(\pi y) |s|}.\end{equation}
This can be expressed in terms of Bessel functions, allowing us to write an explicit form for the large-$N$ limit of Eqs.~\eqref{eq:fkmrawkernels}:
\begin{align}
    K(s) &= 2 \delta(s) + 2 e^{-2|s|} \left[I_0(2|s|)+I_1(2|s|)\right]\;,\nonumber\\
    G(s) &= 2 T  e^{-2|s|} \left[I_0(2|s|)+I_1(2|s|)\right]\;.\label{eq:kernelloop}
\end{align}
where $I_n(z)$ is the $n$th modified Bessel function of the first kind. We conclude from Equation~\eqref{eq:kernelloop} that, even in the large-$N$ limit, the loop model does not satisfy the FDT: unlike in the fully-connected model, the tracer remains out of equilibrium. 

Another distinguishing feature of this model concerns the large $|s|$ tails of the friction and noise kernels, which take the asymptotic form
\begin{equation}K(s) \sim T^{-1}G(s) \sim \frac{2}{\sqrt{\pi |s|}}, \qquad s \gg 1.\end{equation}
This power-law decay in $|s|$, which arises from the asymptotic
scaling $I_n(2 |s|)\sim e^{2|s|}/\sqrt{|s|}$~\cite{arfken2011mathematical}, implies that the loop
model has long-time memory, in contrast to the fully-connected model
for which the kernels have exponential tails.

In this section, we have seen how the generalized Langevin Eq.~\eqref{eq:genlan} allows us to determine the conditions for the equilibration of the tracer in systems where the particles are connected linearly as in Eq.~\eqref{eq:genlanham}. For the loop model, we found that the tracer remains out of equilibrium for all $N$, while for the fully-connected model we demonstrated a large-$N$ equilibrium regime. The latter is particularly interesting since it reproduces results observed in simulations of more realistic baths with short-range interactions~\cite{letter}. We now study in more detail the departure from this equilibrium limit and the emergence of nonequilibrium signatures at large but finite $N$.

\section{Departure from equilibrium in the fully-connected model} \label{sec:active}
We consider the fully-connected model of Fig.~\ref{fig:examples}(a) with a smooth external potential $U$ acting on the tracer. For the sake of generality, we also allow for pairwise interactions between bath particles through a potential $V_b$. We work in one spatial dimension, which allows us to construct a large-$N$ perturbation theory that can be solved to arbitrary order. In Appendix~\ref{appen:ddimperturb}, we generalize our main results to $d$ dimensions. 
\begin{subequations}
\label{eq:eom}
        \begin{eqnarray}
        \dot{x} &=& - \mu U'(x) - \mu \sum_{i=1}^N k(x - x_i)
        \;, \label{eq:eomx}\\* 
        \dot{x}_{i} &=& k(x - x_i)\!-\!\sum_{j=1}^N V_b'(x_i - x_j) + \sqrt{2 T} \eta_i(t)
        \;.
\end{eqnarray}
\end{subequations}
Differentiating Eq.~\eqref{eq:eomx} with respect to time leads to:
\begin{equation}
\ddot{x} = - \mu U''(x)\dot{x} - \mu k N \dot{x} + \mu k \sum_{i} \dot{x}_i
\;.
\end{equation}
All dependence on the bath variables thus occurs through the quantity $\sum_i \dot{x}_i$, which describes the velocity of the bath's center of mass. Since this cannot be influenced by internal forces like $V_b$, we see that the generalized Langevin equation obtained previously by setting $V_b = 0$ is in fact valid for any $V_b$. Using Eqs.~\eqref{eq:eom} to eliminate $\sum_i \dot{x}_i$ in favor of $x$ and $\dot{x}$, we find
\begin{subequations}\label{eq:xdotpdot} \begin{align}
    \ddot{x} &= -\mu kU'(x) - \dot{x} \left[\mu U''(x)+\mu Nk + k\right] + \mu k \sqrt{2TN}\eta(t) .
\end{align}\end{subequations}
We have thus reduced the $(N+1)$ equations~\eqref{eq:eom} into a closed equation for the evolution of $(x,\dot{x})$. We know from Sec.~\ref{subsec:genLanMF} that the $N\rightarrow \infty$ limit of the above is an equilibrium dynamics for the tracer. In this section, we characterize the departure from equilibrium at large but finite $N$ by constructing a perturbation theory in powers of $N^{-1}$.

We note that $\mu$ may be set to unity in Eq. \eqref{eq:xdotpdot} without loss of generality, as this is equivalent to replacing $U(x) \rightarrow~U(x)/\mu$, $T\rightarrow T/\mu$ and $N\rightarrow N/\mu$. We thus hereafter set $\mu = 1$.

\subsection{Perturbative computation of the stationary measure}
We define a rescaled velocity variable $p\equiv \dot{x}/\sqrt{k}$  
as well as a small parameter $\eps \equiv N^{-1}$.
The Fokker-Planck equation for the joint probability density of $(x, p)$ is then $\partial_t \Psi(x,{p}) = \cL \Psi(x,{p})$, where
\begin{align} \label{eq:fokkerplanck}
\frac{\eps}{k} \cL = \frac{\eps}{\sqrt{k}} \left(U' \partial_p - p \partial_x\right) + \left[1 \!+\! \eps\left(\frac{U''}{k}\! +\! 1\right)\right]\partial_p p + T \partial_p^2. 
\end{align}
Our objective is to solve the steady-state equation $\cL \Psi = 0$ perturbatively in powers of $\eps$. To this end, we first write $\Psi(x,p) \propto \exp\left[-\cH(x,p)/T\right]$, where $\mathcal{H}(x,p)$ is an undetermined effective Hamiltonian. The steady-state equation then implies
\begin{align} \label{eq:hammypde} \big\{\eps k^{-1/2} (U' \partial_p -& p \partial_x) + \left[1\!+ \!\eps\left(k^{-1} U'' \!+\! 1\right)\right] p \partial_p  + T \partial_p^2 \big\}\cH
\nonumber\\
= &\left(\partial_p\cH\right)^2 + \eps T \left(k^{-1} U'' \!+ \!1\right)+T.
\end{align}
We now expand $\cH$ as a series in $\eps$:
\begin{equation}\cH = \sum_{n=0}^\infty \eps^n H_n(x,p).
\label{eq:effectiveHamExpansion}\end{equation}
Substituting into~\eqref{eq:hammypde} and matching orders of $\eps$, the leading order equation is $\left(T \partial_p^2 + p \partial_p \right)H_0 = T + \left(\partial_p H_0\right)^2,$
which admits the normalizable solution
\begin{equation}H_0(x,p) = \frac{p^2}{2} + f_0(x),\end{equation}
with $f_0(x)$ undetermined at this order. We thus see that positions and momenta are uncorrelated for $\eps = 0$, as for the equilibrium Maxwell-Boltzmann distribution. We know from Section~\ref{subsec:genLanMF} that, as $N\rightarrow \infty$, an FDT emerges and $x$ becomes Boltzmann-distributed in the potential $U$. We thus make the choice $f_0(x) = U(x)$, which will prove self-consistent from the higher-order equations. The $\cO(\eps)$ equation can then be solved to yield
\begin{equation}H_1 = \frac{1}{2}(k^{-1}U''+1)p^2 + f_1(x),\label{eq:H1sol}\end{equation}
with $f_1(x)$ again undetermined at this order. We see that positions and momenta become correlated when $N$ is finite, a feature typical of active particles \cite{szamel2014self,szamel2015glassy,fodor2016, caprini2020, henkes2020dense}. For $n \geq 2$, the $\cO(\eps^n)$ equation reads
\begin{align} \left(T \partial_p^2\! -\! p \partial_p \right)\!H_n +& \left[\frac{1}{\sqrt{k}}(U' \partial_p \!- \!p \partial_x) \!+\! \left(\frac{U''}{k}\!+\!1\right)\!p \partial_p \right]\!H_{n-1} \nonumber\\
&= \sum_{m=1}^{n-1} \partial_p H_m \partial_p H_{n-m}. \label{eq:hammySeriesPDE}\end{align}
We now further expand $H_n(x,p)$ as a power series in $p$
\begin{equation}H_n(x,p) \equiv \sum_{m=0}^\infty h^{(n)}_m(x) p^m. \label{eq:polyExpansion}\end{equation}
The expansion coefficients $\{\hnp{n}{m}\}$ then obey the following recursive system of ordinary differential equations for $n \geq 2$:
\begin{widetext}
\begin{multline} T(m+1)(m+2) h_{m+2}^{(n)}-m h_m^{(n)}+k^{-1/2} \left[(m+1) U'  h_{m+1}^{(n-1)}-\partial_x h_{m-1}^{(n-1)}\right] + \left(k^{-1}U''+1\right)m h_m^{(n-1)} 
\\ = \sum_{\ell=0}^{m} (m-\ell+1)(\ell+1) \sum_{q=1}^{n-1} h_{m-\ell+1}^{(q)} h_{\ell+1}^{(n-q)},
\label{eq:coefficientRecursion}
\end{multline}
\end{widetext}
We next note that, as shown in appendix~\ref{appen:polynomialHn}, the normalizability of $\Psi$ requires $H_n$ to be a polynomial in $p$ of degree $n+1$, so that $h^{(n)}_m = 0$ for all $m>n+1$. The nonzero terms of the sum on the RHS of Eq.~\eqref{eq:coefficientRecursion} must therefore satisfy $m-\ell \leq q \leq n-\ell$.

Using Eq.~\eqref{eq:coefficientRecursion} and the solutions for $H_0$ and $H_1$, explicit expressions can be obtained for $\hnp{n}{n+1}$ and $\hnp{n}{n}$ (equations~\ref{eq:hnnp1} and~\ref{eq:hnn}). In general, however, solving for $h^{(n)}_{n-k}$ requires knowledge of $h^{(k)}_0$. It is possible to systematically compute each $h^{(k)}_0$ by writing~\eqref{eq:coefficientRecursion} for $n = (1, \dots, k+2)$ and $m = (1, \dots, n+1)$. This procedure is used in appendix~\ref{appen:hnp} to compute $H_1, H_2, H_3$, and the momentum-dependent part of $H_4$, from which various observables of interest can be calculated, as we soon detail. We note that our derivation assumed that $U(x)$ is infinitely differentiable. It would be interesting to generalize beyond this to, for example, piecewise-linear potentials~\cite{letter}. We furthermore note that the perturbative expansion in Eq.~\eqref{eq:effectiveHamExpansion} may in principle be only asymptotically convergent; though we observed no numerical evidence for divergence of the partial sums, such situations may be treated using techniques such as Borel resummation~\cite{serone2017power}.

\subsection{Marginal distribution of the tracer position}
The marginal distribution of the tracer position $x$ can be computed by integrating out $p$. We express this in terms of an effective potential that we expand in powers of $\eps$:
\begin{align} \label{eq:statMeasure}
    P(x) &\equiv \int dp \Psi(x,p) \propto e^{-U_\mathrm{eff}/T}, \nonumber\\
    U_\mathrm{eff}(x) &\equiv U(x) + \sum_{n=1}^\infty \eps^n U_n(x),
\end{align}
The first few $U_n$ are computed in appendix~\ref{appen:marginal}, and show deviation from the Boltzmann distribution at leading order in $\eps$:
\begin{subequations}\label{eq:marginalEffectivePot}
\begin{align}
    U_1 =& \frac{U'^2}{2 k}+U - \frac{T U''}{2 k}, \label{eq:U1}\\
    U_2 =& \frac{T}{2k^2}U''(2k+U''), \\
    U_3 =& \frac{T}{k}\left[U' U^{(3)}-\frac{TU^{(4)}}{2} - U''\left(1+\frac{4+k}{4k}U''+\frac{U''^2}{3k^2}\right)\right] \nonumber\\
    &-\frac{1}{2 k} \int^x U^{(3)}(z) U'(z)^2 \, dz+c \int^x e^{U(z)/T} \, dz,
\end{align}
\end{subequations}
where $c$ is a boundary-dependent integration constant that vanishes for an infinite system with a confining potential, but not for a finite system $(0,L)$ with periodic boundaries $U^{(n)}(0) = U^{(n)}(L)$:
\begin{equation} \label{eq:cdef} c \equiv \begin{cases}
    0, &\text{(Infinite system)}, \\
    \dfrac{1}{2k} \dfrac{\int_0^L dz\, U^{(3)}(z) U'^2(z)}{\int_0^L dz\, e^{U(z)/T}}, &\text{(Periodic boundaries)}.
\end{cases}\end{equation}

Notable in Eqs.~\ref{eq:marginalEffectivePot} is the fact that the effective potential becomes nonlocal in $U$ at $\cO(\eps^3)$. A consequence of this is {density rectification}: Consider an obstacle modeled by a smooth potential $U(x)$ supported on the interval $(a,b)$. In the presence of confining walls at $x_1 < a$ and $x_2 > b$, the non-local dependence of $P(x)$ on $U(x')$ leads to an imbalance in probability density on the two sides of the obstacle:
\begin{equation} \label{eq:rectification} \frac{P(b)-P(a)}{P(a)} = 
\frac{\eps^3 }{2T k} \int_{a}^{b}dz\, U^{(3)}(z) U'(z)^2 + \mathcal{O}(\eps^4).
\end{equation}
As long as the potential $U$ is asymmetric under
$x\to -x$, the integral in Eq.~\eqref{eq:rectification} is nonvanishing and rectification occurs.  We thus see that the tracer density is rectified by asymmetric obstacles; a behavior typical of active particles that was observed numerically in~\cite{letter} for a more realistic model of a cold tracer in a hot bath with short-ranged interactions.

\subsection{Ratchet currents and irreversibility}
If the confining walls are replaced by periodic boundaries, the asymmetric obstacle generates a steady ratchet current $J \equiv \langle \dot{x}\rangle$, whose leading contribution appears at $\cO(\eps^4)$:
\begin{equation}J = \langle \dot{x} \rangle =  \frac{\eps^4}{k} \frac{L\int_0^L dz\, U^{(3)}(z) U'^2(z)}{2 \int_0^L dz\, e^{\frac{-U(z)}{T}} \int_0^L dz\, e^{\frac{U(z)}{T}}},
\label{eq:current}\end{equation}
where $\langle \dot{x} \rangle$ is computed as described in Appendix~\ref{appen:expec}. The fact that the density rectification is present at $\cO(\eps^3)$ while the current appears only at order $\cO(\eps^4)$ reflects the fact that the effective mobility of the tracer vanishes as $1/N$ for large $N$ (Eq. ~\ref{eq:unrescaledkernels}).

The emergence of a steady-state current~\eqref{eq:current} is a direct manifestation of time-reversal symmetry (TRS) breaking. The degree to which TRS is broken can be quantified by the entropy production, which is the Kullback-Leibler divergence from the probability measure of trajectories of a specified length to
the measure of time-reversed trajectories of the same length. According to Eqs.~\eqref{eq:xdotpdot}, the probability density of any path $\{x(s): s \in [0,t]\}$ can be written as~\cite{onsager1953}
\begin{equation}
\cP\left[\{x(s\leq t)\}\right] \propto e^{-\mathcal{S}(t)},
\end{equation}
where the dynamical action $\mathcal{S}(t)$ is given by,
\begin{equation} \label{eq:action}
\mathcal{S}(t) = \int_0^t ds \, \frac{\left[k^{-1} \dot{p} + U' +\left(k^{-1} U'' + N + 1\right)p\right]^2}{4TN},\end{equation}
and the stochastic integral is to be interpreted in the Stratonovich sense. The steady-state entropy production {rate} is then defined as~\cite{maes1999,seifert2005}:
\begin{equation}\sigma \equiv \! \lim_{t\rightarrow \infty} \frac{1}{t} \left\langle\!\log \frac{\mathcal{P}\left\{x(s\leq t)\right\}}{\mathcal{P}\left\{x^\mathrm{R}(s\leq t)\right\}}\!\right\rangle = \! \lim_{t\rightarrow \infty} \!\frac{1}{t} \left \langle \mathcal{S}^\mathrm{R}(t) - \mathcal{S}(t)\right\rangle,
\label{eq:entropyDef}\end{equation}
where the time-reversed action $\mathcal{S}^\mathrm{R}$ is obtained by replacing $\partial_t \rightarrow -\partial_t$ and $p \rightarrow -p$. We then have
\begin{align} \label{eq:SRminusS}
\mathcal{S}^{\mathrm{R}}(t) - \mathcal{S}(t) =& -\frac{1}{T}\int_0^t ds \left\{ \frac{\eps }{k^2} \dot{p} p U'' + \frac{(1+\eps)}{k} \dot{p}p \right.\nonumber \\
&\left. + \frac{\eps}{k}  p U' U''+ (1+\eps) p U' \right\}.
\end{align}
All but the first term of the integrand are total derivatives and therefore integrate to a finite contribution that is not extensive in $t$. Since these do not contribute to Eq.~\eqref{eq:entropyDef}, we find
\begin{equation}
\sigma = - \frac{\eps}{T k^2} \langle \dot{p} p U''\rangle\;.
\end{equation}
In the steady-state, all expectation values are stationary, so that $\partial_t \langle p^2 U'' \rangle = \left \langle 2p \dot{p} U'' + p^3 U''' \right \rangle = 0$. This leads to
\begin{equation} \label{eq:entropy}
\sigma = \frac{\eps}{2Tk^2} \langle p^3 U'''\rangle = \frac{\eps^3 T}{2k^2}\left\langle U'''(x)^2\right\rangle_0 + \cO(\eps^4).\end{equation}
where $\langle \cdot \rangle_0$ denotes an expectation value with respect to the Boltzmann measure $e^{-U/T}$, and where the final equality follows from the procedure outlined in Appendix~\ref{appen:expec}. Irreversibility thus emerges at order $\cO(\eps^3)$.

We note that the deviation from the Boltzmann weight scales as $1/N$ (Eq.~\ref{eq:marginalEffectivePot}), while the entropy production and ratchet current vanish much faster than this, respectively as $1/N^3$ and $1/N^4$. This implies the existence of an intermediate effective equilibrium regime at large but finite $N$, in which the steady-state distribution is non-Boltzmann, but standard manifestations of TRS breaking cannot be detected.

\subsection{Beyond linear coupling and zero-temperature tracers}

It is natural to ask whether the predictions of our perturbation theory extend beyond the setting of a linearly-coupled zero-temperature tracer. We first argue analytically for the universality of the infinite-density equilibrium limit, under the condition that a generalized Langevin dynamics for the tracer exists. Then, we present numerical data showing that our finite-$N$ perturbative results extend to tracers with non-zero temperature, nonlinear fully-connected models, and models with short-ranged repulsive interactions.

Let us first consider the dynamics of a large, overdamped tracer immersed in a fluid at temperature $T_0$, interacting with a bath of Brownian particles at some different temperature $T$ and density $\rho$. So long as the tracer dynamics occur on a time scale larger than the time between its collisions with bath particles, we expect it to follow a generalized Langevin equation \cite{mori1965, vankampen1986, zwanzig2001, granek2022} of a form similar to Eq.~\eqref{eq:genlan}. The friction and noise kernels may be postulated to take the general form
\begin{subequations}\label{eq:generalgenlan}
\begin{align} 
    K(t) &= 2 \mu^{-1} \delta(t) + f(t)\;, \\
    G(t) &= 2 T_0 \mu^{-1} \delta(t) + T f(t)\;,
\end{align}
\end{subequations}
where $f$ is a function satisfying $\lim_{t\rightarrow \infty} f(t) = 0$, which represents the drag and noise due to the hot Brownian particles. The terms proportional to $\delta(t)$ represent the influence of the surrounding fluid at temperature $T_0$. The form postulated in Eqs.~\eqref{eq:generalgenlan} has the properties that if $f(t) = 0$ (i.e., no hot particles), then the tracer satisfies an overdamped Langevin equation at temperature $T_0$, whereas if $\mu^{-1} = 0$ (no cold fluid), then the tracer satisfies an FDT at temperature $T$ due to the hot Brownian bath. This form can be verified explicitly in two settings: first, for the linear fully-connected model (Eq.~\ref{eq:unrescaledkernels}), for which we found $f(t) = N k e^{-k|t|}$, and second, in the context of adiabatic perturbation theory, which applies when there is a separation of timescales between tracer and bath particle dynamics. In the latter case, $f(t)$ can be computed using projection methods \cite{vankampen1986, mori1965, granek2022, solon2018}. 

\begin{figure}
\vspace{2em}
  \begin{overpic}[scale=0.6, trim={5 5 0 11}, clip]{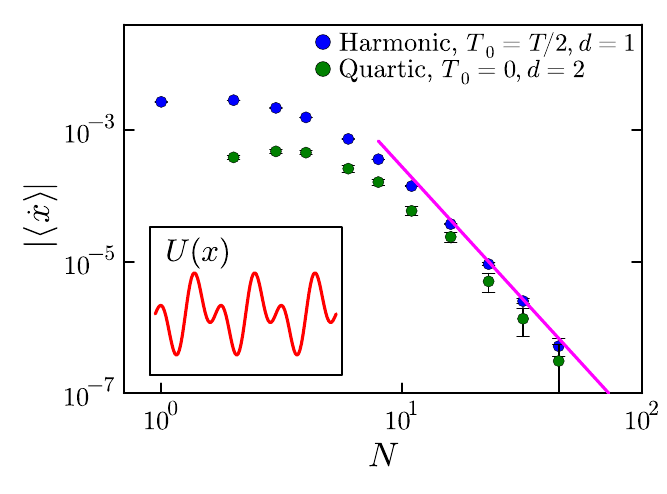} \put(1,67){(b)} 
    \put(72, 37){\small \textcolor{magenta}{$\propto N^{-4}$}}
    \end{overpic}

    \caption{Steady-state current, $\langle \dot{x} \rangle$, in an asymmetric periodic potential $U(x) = \sin(\pi x/2) + \sin(\pi x)$ (inset), for two variants of the fully-connected model: first, a harmonically-coupled model where the tracer temperature is $T_0=T/2$ rather than $T_0 = 0$, and, second, a model in which the harmonic springs coupling tracer and bath particles are replaced by quartic springs with interaction potential $(\br - \br_i)^4/4$. The spatial dimension $d$ is indicated. Both models show an $N^{-4}$ decay, consistent with our perturbative results on the harmonic fully-connected model with a zero-temperature tracer.}
    \label{fig:fcCurrent}
\end{figure}

\begin{figure*}
\vspace{2em}
  \begin{overpic}[scale=0.54, trim={0 2 0 10}, clip]{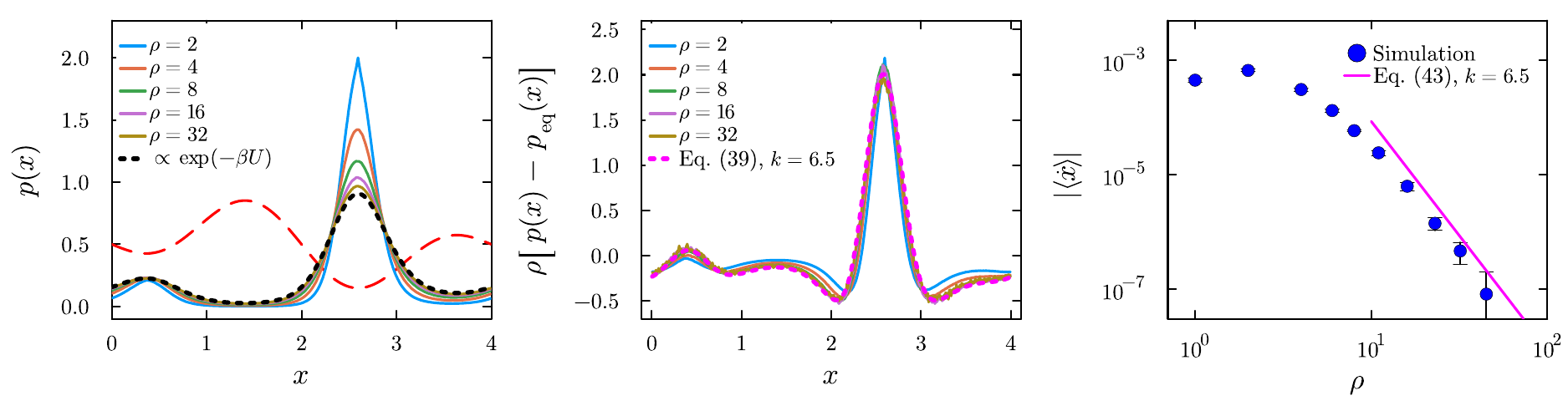}
  \put(0.5,23){(a)} 
  \put(33,23){(b)} 
  \put(67,23){(c)} 

  
  
    \end{overpic}

    \caption{A zero-temperature tracer in a bath of Brownian colloids at temperature $T=0$, with short-ranged soft repulsive interactions between the tracer and bath particles. A periodic potential $U(x) = \sin(\pi x/2) + \sin(\pi x)$ is applied to the tracer. (a) Stationary probability density of the tracer $P(x)$ for different bath densities $\rho$, showing convergence to the Boltzmann weight at large $\rho$. (b) The deviation from the Boltzmann distribution $\left[P(x) - P_\mathrm{eq}(x)\right]$ scales as $1/\rho$, and fits the fully-connected theory of Eq.~\eqref{eq:statMeasure} with an effective $k$ fitted to $k=6.5$. (c) The steady current decays as $1/\rho^4$, with a prefactor comparable to the fully-connected prediction of Eq.~\eqref{eq:current},  with the value $k=6.5$ fitted from panel (b).
    The simulation domain is periodic with size $L=4$ and the interaction potential is $V(r) = 5\Theta(1-r) \times (r-1)^2/2$.
    }
    \label{fig:shortRanged}
\end{figure*}

In general, $f(t)$ is expected to scale with the number of bath particles in contact with the tracer. This applies to the fully-connected model, where we showed $f(t) \propto N$. For models with short-ranged interactions between tracer and bath particles and negligible interactions between bath particles, we expect $f(t) \propto \rho$, where $\rho$ is the density of the bath. This may be verified in adiabatic perturbation theory (see, for example, Refs.~\cite{granek2022, solon2018}). It then follows that, at large $\rho$ (or $N$),  $f(t)$ should dominate the terms proportional to $\delta(t)$ in Eq.~\eqref{eq:generalgenlan}, so that the tracer equilibrates at temperature $T$. We thus predict that the large-density equilibrium limit is a universal feature of cold tracers in hot bath, which holds beyond the linear fully-connected model discussed above.

Having argued for the universality of the  equilibrium limit, we may now ask whether the results of the finite-$N$  perturbation theory of Sec.~\ref{sec:active}A-C apply more generally as well. We begin by considering two variants of the linear fully-connected model: first, a model in which the tracer temperature is increased from zero to $T_0=T/2$ (i.e. half the hot bath temperature), and second, a model where the harmonic coupling between tracer and bath particles is replaced by a quartic one, using $V(\br - \br_i) = c(\br - \br_i)^4/4$. In both cases, numerical simulations in spatial dimensions $d=1,2$ show an $\propto N^{-4}$ scaling of the ratchet current in an asymmetric periodic potential (Fig.~\ref{fig:fcCurrent}), supporting the applicability of our results to more general, nonlinear fully-connected models.

To make contact with experimental systems, such as passive tracers within baths of active enzymes or colloids, it is necessary to move beyond fully-connected models and consider models with short-ranged interactions. To this end, we performed numerical simulations in one spatial dimension of a zero-temperature tracer in a bath of Brownian colloids at temperature $T$ and density $\rho$, with short-ranged soft repulsive interactions. Remarkably, we found that our results on the fully connected model apply to the short-ranged model quite well, with $\rho$ in place of $N$ and a spring stiffness $k$ obtained by fitting. This is illustrated in Fig.~\ref{fig:shortRanged}, which reports simulations of the short-ranged model in an asymmetric periodic potential: the stationary probability density converges to the Boltzmann measure at large $\rho$ (Fig.~\ref{fig:shortRanged}a), and the leading $\mathcal{O}(\rho^{-1})$ correction to Boltzmann matches the fully-connected theory of Eq.~\eqref{eq:statMeasure} upon fitting $k$. Furthermore, the steady current induced by this potential decays as $\rho^{-4}$, hence generalizing directly the $N^{-4}$ scaling of the fully connected case. These results strongly support the universality of our analytical results on the large-$\rho$ scaling of the ratchet current and deviation from the Boltzmann weight, beyond the case of fully-connected models.

\section{Exactly-solvable linear models at the microscopic scale}\label{sec:exact}
So far, we have studied Eq.~\eqref{eq:genEOM} with a quadratic interaction potential $V$ and an arbitrary external potential $U$. If we further specialize to quadratic $U$, then the joint distribution of the tracer and bath particles will be Gaussian and can be computed exactly for any $N$. Although this restriction precludes any TRS-breaking phenomena like ratchet currents, it allows for a non-perturbative characterization of the tracer dynamics and, more importantly, of the effect of the tracer on the bath.

It will prove convenient to combine the tracer and bath degrees of freedom into a single vector $\x = (x, x_1, \dots, x_N)^T \in \mR^{N+1}$, $x \equiv x_0$, so that the total potential energy can be written
\begin{equation}\cH(\x) = \frac{1}{2}\x^T A \x \label{eq:quadham}\end{equation}
where $A$ is a positive-definite symmetric matrix. Equation~\eqref{eq:genEOM} then takes the form
\begin{equation}\dot{\x} = -MA \x + \bm{\Gamma}(t), \label{eq:linearEOM}\end{equation}
where $M$ is the diagonal matrix whose $i$th entry is $\mu_i >0$, and where
\begin{align}
    \langle \Gamma_i \rangle &= 0 \nonumber \\
    \langle \Gamma_i(t) \Gamma_j(t') \rangle &= 2 M_{ij} k_B T \delta(t-t')(1- \delta_{0 i}). \label{noisecorrelator}
\end{align}
Equation~\eqref{eq:linearEOM} defines an $(N+1)$-dimensional Ornstein-Uhlenbeck process, so that $\x$ is a Gaussian random vector characterized completely by its first two moments. The solution to~\eqref{eq:linearEOM} is:
\begin{equation} \label{eomsoln}
\x(t) = G(t) \x(0) + \int_{0}^t ds G(t-s)\bm{\Gamma}(s),
\end{equation}
where $G$ is the (noiseless) Green's function given by:
\begin{equation}\label{eq:greendef} G(t) \equiv e^{-Z t}, \qquad Z \equiv MA.\end{equation}
It can be checked that $Z$ has positive eigenvalues, so that $G$ decays to zero at large times and the stationary dynamics are obtained by neglecting the first term of Eq.~\eqref{eomsoln}. The steady-state, equal-time two-point correlator is then given by:
\begin{equation}\langle x_i x_j \rangle \equiv\lim_{t\rightarrow \infty} \langle x_{i}(t) x_j(t) \rangle = 2T\!\int_{0}^\infty \!\!ds \sum_{k=1}^N G_{ik} G_{jk} \mu_k,  \label{eq:xixj}
\end{equation}
where we emphasize that the sum excludes $k=0$. This is to be compared with the equilibrium result:
\begin{equation}p_{\mathrm{eq}}(\x) \propto \exp \left[-\frac{\x^T A \x}{2T}\right], \qquad \langle x_i x_j \rangle^\eq = T A^{-1}_{ij}.\end{equation}
where $\langle \cdot \rangle^\text{eq}$ denotes an expectation value with respect to the Boltzmann measure $p_\mathrm{eq}$. To measure the deviation from this equilibrium distribution, we introduce the quantity $\Delta_{ij}$ as:
\begin{equation}\Delta_{ij} \equiv T^{-1}\left[\langle x_i x_j \rangle^\eq - \langle x_i x_j \rangle \right]. \label{eq:delDef}\end{equation}
To compute this, we note that
\begin{equation} 
2 \int_0^\infty dt \, G(t) M G^T(t)
= 2\int_0^\infty dt \, e^{-2MAt} M = A^{-1} \label{eq:Ainv}
\end{equation}
where we have used $M e^{-AM t} = e^{-MA t} M$. Equations~\eqref{eq:Ainv} and~\eqref{eq:xixj} then allow us to rewrite Eq.~\eqref{eq:delDef} in terms of $G$ as:
\begin{equation}\Delta_{ij} = 2 \int_0^\infty ds G_{i0}(s) G_{j0}(s)\mu_0 \label{eq:delijInt}\end{equation}
To simplify further, we note that the matrix $Z$, although neither symmetric nor normal in general, is always diagonalizable. This follows from the existence and invertibility of $M^{1/2}$, which allows us to write $M^{-1/2} Z M^{1/2} = M^{1/2} A M^{1/2}$, implying that $Z$ is similar to a symmetric matrix. We may thus expand $Z = P \Lambda P^{-1}$, where $P$ is the matrix whose columns are the right eigenvectors of $Z$ and $\Lambda$ is the diagonal matrix of eigenvalues $\{\lambda_n\}$. The integral in Eq.~\eqref{eq:delijInt} can then be evaluated as:
\begin{equation}\int_0^\infty dt \, G_{i0}(t) G_{j0}(t) = \frac{P_{ik}P^{-1}_{k0}P_{jl}P^{-1}_{l0}}{\lambda_k+\lambda_l},\end{equation}
where summation over repeated indices is implied. This can be written
\begin{equation} \label{eq:momentdev}
    \Delta_{ij} = 2 \mu_0 \sum_{n,\,m=0}^N \frac{v_i^{(n)}u_0^{(n)}v_j^{(m)}u_0^{(m)}}{\lambda_n+\lambda_m},
\end{equation}
where $\{\mathbf{v}^n\}$ and $\{\mathbf{u}^n\}$ are respectively the right and left eigenvectors of $Z$, normalized such that $\bv^{(n)} \cdot \mathbf{u}^{(n)} = 1$. 
Equation~\eqref{eq:momentdev} is a central result of this section: for a given bath topology, it allows us to characterize the departure from thermal equilibrium by diagonalizing the matrix $Z$. 

If $\Delta_{ij} = 0$ for all $i$ and $j$, the zero-temperature reservoir has no effect: the system is in equilibrium at temperature $T$ and the steady-state is Boltzmann-distributed with respect to the Hamiltonian given in Eq.~\eqref{eq:quadham}. When $\Delta_{ij} \neq 0$, the distribution can still be interpreted as an equilibrium distribution \footnote{The Ornstein-Uhlenbeck dynamics~\eqref{eq:linearEOM} always satisfies detailed balance with respect to its steady state distribution.},
but with respect to an effective Hamiltonian given by
\begin{equation}
\cH_\mathrm{eff} = \frac{1}{2}\x \left(A^{-1} - \Delta\right)^{-1} \x.
\end{equation}
We expect that the cold reservoir connected to the tracer suppresses fluctuations within the system. We quantify this by computing the ratio 
\begin{equation}\label{eq:correlRat}
R_{ij} \equiv \frac{\langle x_i x_j \rangle}{\langle x_i x_j \rangle^\eq}  = 1 - \frac{\Delta_{ij}}{A_{ij}^{-1}}\;.
\end{equation}
We now apply the formalism above to characterize the bath and tracer fluctuations in linearized versions of the fully-connected and loop models of Fig. \ref{fig:examples}. In Appendix~\ref{appen:nonlinear}, we show how this approach can be extended perturbatively to weakly non-linear potentials acting on the tracer.

\subsection{Fully-connected model} \label{subsec:linearmf}
The most general model in which the tracer is linearly coupled to $N$ identical, noninteracting bath particles is defined by:
\begin{equation}\cH = \frac{1}{2} k_0 x_0^2 + \sum_{i=1}^N \frac{1}{2}k x_i^2 - c x_0 x_i. \label{eq:mfhammy}\end{equation}
For simplicity, we set all mobilities to unity $M_{ij} = \delta_{ij}$, so that $Z$ is symmetric with coinciding left and right eigenvectors. We then have
\begin{equation}Z_{ij} = k \delta_{ij} - c (\delta_{i0} + \delta_{j0}) + (2c - k + k_0)\delta_{i0}\delta_{j0}.\end{equation}
To diagonalize $Z$, first note that the $N\times N$ block obtained by removing the zeroth row and column of $Z$ is equal to $k \delta_{ij}$. This means that any $\bv \in \mathbb{R}^{N+1}$ with $v_0 = 0$ and $\sum_{i=1}^N v_i = 0$ is an eigenvector of $Z$ with eigenvalue $k$. There are $N-1$ such linearly independent eigenvectors, and they do not contribute to Eq.~\eqref{eq:momentdev} because their zeroth entries vanish. The remaining two eigenvectors can be determined by making an ansatz $v_{i>0} = 1$ and solving the eigenvalue equation for $v_0$. We find two solutions:
\begin{equation}v_0^{\pm} = \frac{1}{2c}\left(k-k_0 \pm \sqrt{(k-k_0)^2+4N c^2}\right),\end{equation}
with corresponding eigenvalues $\lambda_{\pm} = k - c v_0^{\pm}$. After normalizing $\bv$ and substituting into Eq.~\eqref{eq:momentdev}, we obtain
\begin{eqnarray}
   \hspace{-1em} \Delta_{ij} = \frac{1}{(k+k_0)(k k_0 - c^2N)} \times& \hfill \nonumber \\
    & \hspace{-6em} \begin{cases}
        k (k_0+k)-c^2N, &(i=j=0), \\
        c^2, &(i\neq0, j\neq0) \\
        ck, & \text{otherwise}. \\
    \end{cases}
\end{eqnarray}
It is straightforward to verify that 
\begin{equation}A_{0j}^{-1} = \frac{c+(k-c)\delta_{j0}}{k_0 k - c^2 N}, \quad A^{-1}_{pq} = \frac{\delta_{pq}}{k}+\frac{c^2 }{k(k_0k-c^2N)},\end{equation}
for $p,q\neq 0$. To interpret these results, we note that the ratio $R_{ii} =\langle x_i^2\rangle/\langle x_i^2\rangle^{\mathrm{eq}}$ can be thought of as a dimensionless effective temperature for particle $i$, and can be determined from Eq.~\eqref{eq:correlRat} as
\begin{equation}R_{00} = \frac{c^2 N}{k^2 + k k_0}, \quad R_{jj} = 1-\frac{c k^2}{(k+k_0)\left[k k_0-c^2(N-1)\right]},\end{equation}
for $j \neq 0$. To study the large $N$ limits of these results, we must make specific choices for $k$, $c$ and $k_0$ as functions of $N$. We first consider the model depicted in Fig.~\ref{fig:examples}(a) with $V_b = 0$ and $U(x) = ux^2/2$. This corresponds to $c=k$ and $k_0 = u + kN$. We furthermore allow for weak coupling, in which $k$ is made to vanish with $N$ as $k \sim \tilde{k} N^{-\alpha}$ for some $\alpha>0$. We then find
\begin{subequations} \label{eq:linearRijSpring}
\begin{eqnarray} \label{eq:linearRijSpringA} \lim_{N\rightarrow \infty} R_{00} &&= \begin{cases}
    1, &\alpha < 1, \\
    \dfrac{\tilde{k}}{\tilde{k}+u}, &\alpha = 1, \\
    0, &\alpha > 1.
\end{cases} \\ \label{eq:linearRijSpringB}
\lim_{N\rightarrow \infty} R_{jj} &&= 1,  \qquad j>0.\end{eqnarray}
\end{subequations}
Eq.~\eqref{eq:linearRijSpringA} confirms the result of Sec.~\ref{sec:genlan} that the tracer equilibrates at temperature $T$ for $\alpha < 1$ and follows a zero-temperature dynamics for $\alpha > 1$. For the marginal case $\alpha = 1$, the linearity of the model forces the tracer to equilibrate at some temperature, but the result above tells us that this temperature can be tuned between zero and $T$ by varying $\tilde{k}$ and $u$. Eq.~\eqref{eq:linearRijSpringB} shows that in such a fully-connected model, the bath variables remain equilibrated at temperature $T$ for $N\gg1$, regardless of the status of the tracer.

For completeness, we mention an alternative choice for the parameters $c$ and $k_0$: In their study of the thermalization of Hamiltonian dynamical systems, Caldeira and Leggett considered an interaction similar to Eq.~\eqref{eq:mfhammy}, but with an arbitrary potential on the tracer and an additional so-called ``counter-term" of the form $c^2 N x_0^2/2k$ that keeps the system mechanically stable as $N\rightarrow \infty$. If we choose a quadratic external potential $u x_0^2/2$, then this corresponds exactly to Eq.~\eqref{eq:mfhammy} but with $k_0 = u + c^2N/k$. Allowing for weak coupling $c \sim \tilde{c} N^{-\alpha}$, we find
\begin{subequations}
\begin{eqnarray} \lim_{N\rightarrow \infty} R_{00} &&= \begin{cases}
    1, &\alpha < 1/2, \\
    \dfrac{\tilde{c}^2}{\tilde{c}^2+ k(k+u )}, &\alpha = 1/2, \\
    0, &\alpha > 1/2.
\end{cases} 
\\
\lim_{N\rightarrow \infty} R_{jj} &&= 1,  \qquad j>0.\end{eqnarray}\end{subequations}
The Caldeira-Leggett model thus exhibits a similar transition as $\alpha$ is increased, with equilibrium dynamics at temperature $T$ for $\alpha<1/2$, a tunable temperature for $\alpha=1/2$, and zero-temperature behavior for $\alpha>1/2$.

\subsection{The loop model} \label{sec:loopmodelexact}
We now return to the loop model of Fig.~\ref{fig:examples}(b). We first confirm that the model never equilibrates and then characterize the effect of the tracer on the bath. Interestingly, we find the latter to be long-ranged, leading to a suppression of fluctuations in the bath that decays as a power law with the distance to the tracer.

Setting $U=0$ for simplicity, the interaction energy is
\begin{equation}\label{eq:loopHam}
\cH = \frac{1}{2}\sum_{i=0}^N (x_{i+1} - x_i)^2,\end{equation}
where $x_{N+1} \equiv x_0$ and the spring stiffness has been set to unity without loss of generality. We furthermore set $\mu_i~=~1$ for all $i$, so that the matrix $Z \equiv MA$ is given by
\begin{equation}Z_{ij} = 2\delta_{ij} - \delta_{|i-j|,1} -\left(\delta_{i0} \delta_{jN}+ \delta_{iN} \delta_{j0}\right).\end{equation}
This is a symmetric circulant matrix that is diagonalized in a discrete Fourier basis. The orthonormal eigenvectors are, for $k \in (0, \dots, N)$,
\begin{equation}v_j^{(k)}=\frac{w^{jk}}{\sqrt{N+1}}\end{equation}
where $w$ is the $(N+1)$st root of unity
\begin{equation}w \equiv \exp \left(\frac{2 \pi i}{N+1}\right).\end{equation}
The corresponding eigenvalues are
\begin{equation}\lambda_k = 2 - w^k - w^{Nk}.\end{equation}
There is a zero eigenvalue $\lambda_0$ corresponding to global translations, which dominate the fluctuations of $x_i$.
\if{To make $A$ positive-definite, we introduce a regularizer $Z \rightarrow Z + \eps \mathbbm{1}$. This leaves the eigenvectors unchanged at $\mathcal{O}(\eps)$, but shifts the eigenvalues $\lambda_k \rightarrow \lambda_k + \eps +\mathcal{O}(\eps^2)$. Equation~\eqref{eq:momentdev} then yields
\begin{equation}\Delta_{ij} = \frac{2}{(N\!+\!1)^2} \!\sum_{k,\ell=0}^N \frac{w^{i k + j \ell}}{4\! +\! 2 \eps - w^k \!- w^\ell \!- w^{N k} \!- w^{N \ell}}.\end{equation}

The $k = \ell = 0$ term in the sum above diverges as $\eps \rightarrow 0$, reflecting the fact that the correlator $\langle x_i x_j \rangle$ is dominated by center-of-mass motion.}\fi
To circumvent this, we change to relative displacements by defining $\tilde{\Delta}_{i\neq j}$ as
\begin{align}
    \tilde{\Delta}_{ij} &\equiv  T^{-1} \left[\left\langle (x_i -x_j)^2 \right\rangle^\eq -\left \langle (x_i -x_j)^2 \right \rangle\right] \nonumber \\
    &= \Delta_{ii} + \Delta_{jj} - 2 \Delta_{ij} \nonumber\\
    &= \frac{2}{(N\!+\!1)^2} \!\sum_{k,\ell=0}^N \frac{w^{i(k+\ell)}+w^{j(k+\ell)}-2w^{i k + j \ell}}{4 - w^k - w^\ell - w^{Nk} - w^{N \ell}}\;. \label{eq:deltildedisc}
\end{align}

Introducing the continuous variables $x \equiv 2\pi k/N, y \equiv 2\pi \ell/N$, the large $N$ limit of $\tilde \Delta$ can be expressed as an integral:
\begin{equation} \label{eq:deltilde}
    \lim_{N\rightarrow \infty}\! \tilde{\Delta}_{nm} =\!\int_0^{2\pi}\!\!\!\int_0^{2\pi} \!\frac{dx dy}{(2\pi)^2} \, \, \frac{\left(e^{i n x}-e^{i m x}\right) \left(e^{i n y}-e^{i m y}\right)}{2 - \cos(x) - \cos(y)},
\end{equation}
where we have explicitly symmetrized in $n$ and $m$ so that the integrand is analytic at the origin. Note that, in taking the infinite loop limit to go from Eq.~\eqref{eq:deltildedisc} to Eq.~\eqref{eq:deltilde}, we lose periodicity under $n\to n + N+1$. Equation~\eqref{eq:deltilde} is thus quantifying the fluctuations between two points on the same side of the loop at finite distances $n$ and $m$ from the tracer. 

It is instructive to consider $\tilde{\Delta}_{m, m+1}$, which measures the deviation of the mean squared length of the $m$th spring from its equilibrium value, normalized by temperature. This can be written, for large $N$, as
\begin{equation} \label{eq:kkplus}
    \tilde{\Delta}_{m, m+1} =\int_{0}^\pi \frac{dz}{\pi^2}   \cos\left[(2m\!-\!1)z\right]\left[2z \sec(z)-\pi \tan(z)\right],
\end{equation}
which follows from Eq.~\eqref{eq:deltilde} after making the variable change $2z \equiv x+y, 2w \equiv x-y$, integrating over $w$, and using the integrality of $m$ to simplify the result. It can be furthermore shown that, for integer $m$, the integral in Eq.~\eqref{eq:kkplus} is equal to
\begin{align} \label{eq:kkplusExact}
\tilde{\Delta}_{m, m+1} = \frac{1}{\pi}&\left[\frac{2}{2m+1}\right.+ 2\pi(-1)^m  \nonumber\\ 
&+\psi\left.\left(\frac{1}{4}-\frac{m}{2}\right)-\psi\left(-\frac{1}{4}-\frac{m}{2}\right)\right],
\end{align}
where $\psi$ denotes the digamma function. We note that we can extend Eq.~\eqref{eq:kkplusExact} to negative values of $m$ to describe particles on the left of the tracer, using $\tilde{\Delta}_{-m-1, -m}\equiv \tilde{\Delta}_{m, m+1}$. 
To quantify the degree to which the cold tracer suppresses the fluctuations of the bath, we may assign an effective temperature to the $m$th particle by averaging the fluctuations of its two neighboring springs
\begin{equation}T_\mathrm{eff}(m) \equiv \frac{1}{2} \left\langle(x_{m} - x_{m-1})^2 + (x_{m+1} - x_m)^2\right\rangle. \label{eq:TeffDef} \end{equation}
In equilibrium, the above is equal to $T$ for all $m$ by the equipartition theorem. Adding the cold tracer at $m=0$ will reduce $T_{\mathrm{eff}}(m)$ from its equilibrium value in accord with Eq.~\eqref{eq:kkplusExact}. Defining $\Delta T(m) = T - T_{\mathrm{eff}}(m)$, we find, using the identity $\psi(z+1)=\psi(z)+z^{-1}$, that
\begin{equation}\label{eq:deltaT}
\Delta T(0) = \frac{\pi -2}{\pi} T, \qquad \Delta T(m\neq 0) = \frac{2 T}{\pi} \frac{1}{4m^2-1} \;.
\end{equation}
We see from Eq.~\eqref{eq:deltaT} that the effect of the cold tracer decays as $m^{-2}$ for $m\gg 1$ and is therefore long-ranged. This means that the single cold particle cools down the entire loop. 

Equations~\eqref{eq:kkplusExact} and~\eqref{eq:deltaT} are exact for the one-dimensional chain depicted in Fig.~\ref{fig:examples}(b). In section~\ref{sec:fieldtheory}, we present an alternative, field-theoretic derivation of this $m^{-2}$ scaling that allows generalization to higher-coordination lattices.

\section{Field theory of a cold particle in a hot lattice}\label{sec:fieldtheory}
In this section, we consider a generalized version of the loop model of Fig.~\ref{fig:examples}(b), in which the one-dimensional chain is replaced by a $d$-dimensional hypercubic lattice. For generality, the particle coordinates $\{\vec{x}_i\}$ are allowed to be $n$-dimensional, which allows for a study of spring networks that are confined to lower dimensions $(n<d)$, or embedded in higher-dimensional spaces $(n>d)$--- think about a two-dimensional (2D) sheet confined to a 1D tube or embedded in a 3D space. We first consider a microscopic model that can be coarse-grained exactly for direct comparison to the predictions of section~\ref{sec:exact}, and then study a more general continuum elastic theory.

\subsection{From micro to macro: an exactly solvable case} \label{sec:coarseGrainedLattice}
We index the particles by $\mathbf{m} \in \mathbb{Z}^d$, where $m_k$ labels the particle's coordinate along the lattice direction $k \in 1, \dots, d$. The corresponding particle position is denoted $\vec x_{\mathbf{m}}$. The number of particles along each lattice dimension is denoted $L$, so that $N = L^d$. 
The most direct generalization of the loop model, as defined by Eq.~\eqref{eq:loopHam}, is to let the $\mathbf{m}$th particle evolve according to
\begin{equation}\dot{\vec{x}}_\mathbf{m}= \sum_{\mathbf{p} \in \sigma(\mathbf{m})} (\vec{x}_\mathbf{p} - \vec{x}_\mathbf{m}) + \sqrt{2T}\vec{\eta}_\mathbf{m}(t),
\label{eq:zeroRestLengthLangevin}\end{equation}
where $\sigma(\mathbf{m})$ denotes the nearest neighbors of $\mathbf{m}$, and $\vec{\eta}_\mathbf{m} \in \mathbb{R}^n$ has correlations
\begin{equation}\langle \eta^\alpha_\mathbf{m}(t) \eta^\beta_\mathbf{p}(t) \rangle = \delta_{\alpha \beta}\delta_{\mathbf{m}, \mathbf{p}}(1-\delta_{\mathbf{m}, \mathbf{0}}) \delta(t - t')\;.\label{eq:discnoisecor}\end{equation}
The noise $\eta_\mathbf{m}(t)$ thus acts on all particles except the cold tracer at $\vec{x}_\mathbf{0}$.

To coarse-grain this model into a field theory, we define a continuous variable $\br \in \mathbb{R}^d$ and a displacement field $\vec{u}(\mathbf{r}) \in \mathbb{R}^n$ via
\begin{equation}
r_k \equiv L^{-1} m_k, \qquad \vec{u}\left(L^{-1} \mathbf{m}\right) \equiv L^{\frac{d-2}{2}} \vec{x}_{\mathbf{m}}. \label{eq:fielddef}
\end{equation}
The choice to scale $u$ as $L^{\frac{d-2}{2}}$ will prove to keep $u\in \mathcal{O}(1)$ as $L$ is made large. Upon a diffusive rescaling of time $t \rightarrow t/L^2$, the dynamics of $\vec{u}(\br, t)$ read,
\begin{align}\partial_t \vec{u} =& L^2 \sum_{i=1}^d \big[\vec{u}(\br + L^{-1} \hat{\mathbf{e}}_i) + \vec{u}(\br - L^{-1} \hat{\mathbf{e}}_i) -2\vec{u}(\br)\big] \nonumber\\
&+ \sqrt{2T L^{d}} \vec{\eta}_{L\br}(t),\end{align}
where $\hat{\mathbf{e}}_i$ is the unit vector along lattice direction $i$. Taking $L$ large, we find
\begin{equation}\partial_t \vec{u}(\br, t) = \nabla^2 \vec{u} + \vec{\Lambda}(\br, t), \label{eq:dtu_zerorestlength}\end{equation}
where the correlations of $\vec{\Lambda}$ are obtained from Eq.~\eqref{eq:discnoisecor} using $\lim_{L\rightarrow \infty} L^d \delta_{L\br, L\br'} = \delta(\br - \br')$:
\begin{equation}\langle \Lambda_i(\br, t) \Lambda_j(\br', t') \rangle = 2T \delta_{ij}\delta(t-t') \delta(\br - \br')\left(1 - L^{-d} \delta(\br)\right).
\label{eq:lambdacor}\end{equation}
In Fourier space, Eq.~\ref{eq:dtu_zerorestlength} reads $\partial_t \vec{u}(\q,t) = -q^2 \vec{u} + \vec{\Lambda}(\q,t)$,
where $\vec{u}(\q,t) \equiv \int d^{d}\br e^{-i \q \cdot \br} \vec{u}(\br, t)$, and the Fourier-space noise correlations can be shown to equal
\begin{equation}\langle \Lambda_i(\q,t) \Lambda_j(\q',t') \rangle = 2 T\delta_{ij}\delta(t-t')\!\left[(2\pi)^d \delta(\q\!+\!\q')\!-\!L^{-d}\right].\end{equation}
The long-time solution is 
\begin{equation}\vec{u}(\q,t) = \int_{-\infty}^t ds e^{-q^2(t-s)} \vec{\Lambda}(\q, s),\end{equation} 
from which it follows that
\begin{equation}
    \langle u_i(\q,t) u_j(\q',t) \rangle = T \delta_{ij} \left[(2 \pi)^d \frac{\delta(\q + \q')}{q^2} - \frac{2 L^{-d}}{q^2+q'^2}\right]. \label{eq:ftFieldcorr}
\end{equation}

We are here interested in the fluctuations of the springs connecting the particles and not in the absolute displacements, leading us to consider
\begin{align}\langle \nabla u_i(\br,t) \cdot \nabla u_j(\br',t)\rangle& \nonumber
\\= -\int \frac{d^d \q d^d \q'}{(2\pi)^{2d}}& e^{i \q \cdot \br + i \q'\cdot\br'} \q\cdot \q' \langle u_i(\q,t) u_j(\q',t) \rangle \label{eq:inverseFT}
\end{align}
The first term of Eq.~\eqref{eq:ftFieldcorr} contributes a term $T \delta(\br - \br')$ to the above. The contribution of the second term involves the integral
\begin{equation}\int \frac{d^d \q d^d \q'}{(2\pi)^{2d}} \frac{e^{i \q \cdot \br + i \q'\cdot\br'} \q\cdot \q'}{q^2+q'^2} = \nabla_\br \cdot \nabla_{\br'} \int \frac{d^{2d} \tilde{\q}}{(2\pi)^{2d}} \frac{e^{i \tilde{\q}\cdot \tilde{\br}}}{\tilde{q}^2},\end{equation}
where $\tilde{\br}^T \equiv (\br^T, \br'^T)$, and similarly for $\tilde{\q}$. The integral on the right hand side is a $2d$-dimensional Coulomb potential \cite{kardar2007} evaluating to
\begin{equation}\int \frac{d^{2d} \tilde{\q}}{(2\pi)^{2d}} \frac{e^{i \tilde{\q}\cdot \tilde{\br}}}{\tilde{q}^2} = \begin{cases}
    -\log(r^2+r'^2)/4\pi, &d = 1, \\
    \frac{(d-2)!}{4\pi^d}(r^2+r'^2)^{1-d}, &d \geq 2.
\end{cases}\end{equation}
It then follows that, for all $d$,
\begin{align}\langle \nabla u_i(\br,t) \cdot \nabla u_j&(\br',t)\rangle \nonumber \\= T\delta_{ij} &\left[\delta(\br - \br') - \frac{2 d!}{\pi^d L^d} \frac{\br \cdot \br'}{(r^2+r'^2)^{1+d}}\right]. \label{eq:twoPointfield}\end{align}
The first term above is the equilibrium result that would be obtained if all particles were at temperature $T$. The second term represents the suppression of fluctuations due to the cold particle at $r=0$, which decays as $r^{-2d}$ at large distances. 

We can connect back to the microscopic model by generalizing the definition in Eq.~\eqref{eq:TeffDef} to arbitrary $d$ and $n$:
\begin{equation}
    T_\mathrm{eff}(\mathbf{m}) \equiv \frac{1}{2n}\sum_{\mathbf{p} \in \sigma(\mathbf{m})} \left \langle(\vec{x}_\mathbf{p} - \vec{x}_\mathbf{m})^2 \right \rangle\;.
\end{equation}
From the definitions of Eq.~\eqref{eq:fielddef}, we have
\begin{align}
    T_\mathrm{eff}(\mathbf{m})\!= \!\!\! \!\sum_{\mathbf{p} \in \sigma(\mathbf{m})} \!\!\!\! \frac{\big\langle\!\left[ \vec{u}(L^{-1}\mathbf{p}) - \vec{u}(L^{-1}\mathbf{m})\right]^2\!\big \rangle}{2n L^{d-2}} = \frac{\big\langle \!\left[\nabla \vec{u}(\br,t)\right]^2 \!\big\rangle}{n L^d}
\end{align}
Using $\delta(0) = L^{d}$, it can then be shown that the temperature suppression satisfies 
\begin{equation}
    \Delta T(\mathbf{m}) \equiv T - T_\mathrm{eff}(\mathbf{m}) = \frac{d!}{(2\pi)^d}\frac{T}{m^{2d}}. \label{eq:tefffieldtheory}
\end{equation}
Note that for $d = n = 1$, this is identical to the result of the exact calculation, Eq.~\eqref{eq:deltaT}, for $m\gg1$. We conclude that the effect of a cold particle within a hot, $d$-dimensional hypercubic lattice decays with the distance to the tracer as $m^{-2d}$. 
\label{sec:zeroRestLength}
\subsection{Phenomenological theory of a cold inclusion in a hot elastic medium} \label{sec:elasticTheory}
The model defined in Eq.~\eqref{eq:zeroRestLengthLangevin} represents a lattice of points (or displacements) $\vec x_{\mathbf{m}}$ connected by springs of zero rest length.
It does not resist shear or compression at zero temperature and thus does not reflect the elastic properties of most realistic materials. 
To extend our results to such materials, we use a standard phenomenological linear elastic theory. 
We let $\bu(\br, t)$ denote the displacement of a material particle from its equilibrium position at $\br$ and consider the Hamiltonian~\cite{landauElasticity1986}
\begin{equation}\label{eq:HamLandau}
\mathcal{H} = \int d^d\br \left[\mu (u_{ik})^2 + \frac{1}{2}\lambda (u_{ii})^2\right]\;,
\end{equation}
where $u_{ij}(\br) \equiv (\partial_{i}u_j + \partial_j u_i)/2$ is the symmetrized strain tensor, and $\mu$ and $\lambda$ are the two Lamé coefficients. To describe the evolution of a gel-like material (or of a solid coupled to a substrate~\cite{scheibner2020odd}) we use an overdamped dynamics for the lattice
\begin{equation}\partial_t \bu(\br, t) = -\frac{\delta \mathcal{H}}{\delta \bu(\br)} + \blambda(\br, t)\;,
\end{equation}
where we have chosen the units of time such that the  field has unit mobility, and the correlations of $\blambda(\br, t)$ are given by Eq.~\eqref{eq:lambdacor} as before. We  define $\kappa \equiv (\mu + \lambda)/\mu$ and  set $\mu = 1$, which is equivalent to a rescaling of time and temperature as $t \rightarrow \mu t, T \rightarrow T/\mu$. The dynamics of $\bu$ then read
\begin{equation}\partial_t \bu(\br,t) = \nabla^2 \bu + \kappa \nabla(\nabla \cdot \bu) + \blambda(\br, t)\;.
\label{eq:elasticfieldTheory}\end{equation}
The second term of Eq.~\eqref{eq:elasticfieldTheory}, although allowed by symmetry, did not appear in the coarse-grained description (Eq.~\ref{eq:dtu_zerorestlength}) of the simpler model defined in Eq.~\eqref{eq:zeroRestLengthLangevin}. The solution to Eq.~\eqref{eq:elasticfieldTheory} is obtained in Fourier space as
\begin{equation}\bu(\q, t) = \int_{-\infty}^t ds \exp\left[-(q^2 \mathbbm{1} + \kappa \q \otimes \q)(t-s)\right] \blambda(\q, s),\end{equation}
where $(\q \otimes \q)_{ij} \equiv q_i q_j$. To diagonalize the exponential, we split $\blambda(\q, s)$ into longitudinal and transverse components:
\begin{equation}\blambda(\q, t) = \blambda^\parallel(\q, t) + \blambda^\perp (\q,t).\end{equation}
where $\blambda^\parallel \equiv (\hat{\q} \cdot \blambda)\hat{\q}$ and $\blambda^\perp \equiv \blambda - \blambda^\parallel$. The former is an eigenvector of $\q \otimes \q$ with eigenvalue $q^2$, while the latter has eigenvalue zero. We thus have
\begin{equation}\bu = \int_{-\infty}^t ds e^{-q^2(t-s)}\left[e^{-\kappa q^2(t-s)} \blambda^\parallel(\q, s) + \blambda^\perp(\q,s) \right]. \label{eq:elasticSol}\end{equation}
The two-point correlator $\left\langle u_i(\q,t) u_j(\q',t)\right\rangle$ is computed in Appendix~\ref{appen:elastic}. A simple scaling analysis then shows that the influence of the cold tracer on the fluctuations of the strain tensor should decay as $1/r^{2d}$, similar to the $\kappa = 0$ case studied in Sec.~\ref{sec:coarseGrainedLattice}. To demonstrate this, we compute the effect of the cold tracer on local compressions and expansions of the material, which are quantified by the trace of the strain tensor $u_{ii}= \nabla \cdot \bu$. As shown in Appendix~\ref{appen:elastic}, this leads to a form very similar to Eq.~\eqref{eq:twoPointfield}:
\begin{align}\langle (\nabla \cdot \bu)(\br,t)& (\nabla \cdot \bu)(\br',t)\rangle \nonumber \\= \frac{T}{1+\kappa} &\left[\delta(\br - \br') - \frac{2 d!}{\pi^d L^d} \frac{\br \cdot \br'}{(r^2+r'^2)^{1+d}}\right]. \label{eq:twoPointElastic}\end{align}
The effect of the cold tracer is thus long-ranged and decays as $r^{-2d}$, as before.

We note that the long-ranged effect of the cold inclusion is a purely nonequilibrium phenomenon. Indeed, if instead of cooling the inclusion at the origin we impose a localized strain through an external potential, the effect is local: As shown in Appendix~\ref{appen:elastic}, the last term in Eq.~\eqref{eq:twoPointElastic} proportional to ${\br \cdot \br'}/{(r^2+r'^2)^{1+d}}$ is then replaced by one proportional to $\delta (\br-\br')$.

Finally, it might be tempting to predict the local effective temperature of the gel using the heat equation with a cold sink at the origin. We note, however, that the presence of hot and cold heat reservoirs makes energy a non-conserved field, so that drawing predictions from the heat equation is not justified
~\cite{bonetto2000fourier}. Indeed, it is easy to check that $T_{\rm eff}(\br)=T-a r^{-2 d}$ is not a solution to $\nabla^2 T_{\rm eff}(\br)=0$. 

\section{Conclusion}
In this work, we have developed a general theory of the dynamics of a cold tracer particle within a bath of linearly-coupled hot Brownian particles. 
We first constructed the generalized Langevin equation of the tracer for the most general linearly-coupled bath, allowing for an arbitrary external potential acting on the tracer (Sec.~\ref{sec:genlan}). 
We applied this result to two specific models: first, a fully-connected model in which the tracer interacts harmonically with $N$ hot particles at once, and second, a model in which the cold tracer is inserted within a harmonic loop of hot particles. 
The former can be seen as a mean-field model of a tracer in a fluid \cite{letter}, and was found to transition from an FDT-violating dynamics at finite $N$ to an equilibrium dynamics at large $N$. 
The loop model, which can be thought to describe a gel, was found to violate the FDT irrespectively of the size of the loop. 
Our results thus suggest that a cold tracer in a fluid of hot colloids will equilibrate only if the density of colloids or radius of the tracer is sufficiently large, while a cold particle in a hot gel will never equilibrate.

In Sec.~\ref{sec:active}, we studied the emergence of nonequilibrium features in the fully connected model perturbatively for large but finite $N$. We found that the tracer departs from the Boltzmann distribution at leading order in $N^{-1}$ and, at higher orders, displays signatures of irreversibility such as entropy production, ratchet currents and long-ranged density rectification by asymmetric obstacles. These findings establish analytically a connection between the dynamics of a cold tracer in a hot bath and those of an active particle~\cite{letter, burov2024}.

Prior to this work and its companion Letter~\cite{letter}, existing results on the behavior of a cold tracer in a hot bath pertained only to free or harmonically trapped tracers, focusing on such quantities as the long-time diffusivity and drag coefficients of the tracer~\cite{ilkerjoanny2021,grosberg2021, jardat2022diffusion, goswami2023trapped}. These results suggested the cold tracer to be amenable to an equilibrium description with an effective temperature, in spite of the theoretical possibility of a nonequilibrium dynamics driven by the steady heat flux~\cite{visco2006work,Zia_2007,li2019quantifying}. Our work demonstrates conclusively that the dynamics of the tracer cannot be adequately described by equilibrium physics outside the large-density limit.

To characterize the impact of the tracer on the bath itself, we considered in Sec~\ref{sec:exact} the case of a linear bath coupled to a harmonically trapped tracer, which is exactly solvable. We applied this to study the suppression of bath fluctuations by the cold tracer in the loop model and found it to be long-ranged, decaying with the inverse square of the distance to the tracer. We generalized this result to higher-dimensional lattices and elastic media in Sec~\ref{sec:fieldtheory}, finding that the damping of fluctuations decays with the distance to the cold particle as $r^{-2d}$. A cold inclusion in a hot gel will thus cool down the entire gel.

Beyond their theoretical interest, the models introduced in this article may be of experimental relevance because they describe the dynamics of a passive tracer within any bath whose fluctuations are enhanced by an external energy source. 
The external drive could arise, for example, by irradiation of the bath particles~\cite{volpe2011microswimmers}, or due to their coupling to an out-of-equilibrium chemical reaction as in the case of active enzyme solutions~\cite{ghosh2021, muddana2010substrate, sengupta2013enzyme, dey2015micromotors, zhao2017enhanced}.
For a passive colloid in an active enzyme solution, it can be argued that, thanks to scale separation, a micron-sized tracer will interact with many bath particles at once, so that the large-$N$ regime may be within reach. We note that, when the tracer is embedded in a momentum-conserving fluid, hydrodynamics may play an important role~\cite{Leptos2009,Dunkel2010,Morozov2014,Thiffeault2015}, for instance leading to the emergence of long-time tails~\cite{van1982transport}. We believe that our results should be robust to these effects since our fully-connected and short-ranged models agree very well, despite the latter having an additional conserved density field. 
The existence of additional hydrodynamic modes thus does not seem to significantly impact our predictions.
Beyond the infinite $N$ limit, our predictions on the scaling of the steady current with the bath density could be verified, for example, using asymmetric corrugated substrates. 
As for our results on cold inclusions in hot elastic media, we believe that active solids with locally tunable activity constitute an ideal experimental platform to measure the predicted long-ranged correlations~\cite{massana2024multiple, baconnier2022selective}.


\noindent\textbf{\textit{Acknowledgements.}}
We thank Hugues Chaté, Omer Granek, Yariv Kafri, Mehran Kardar, Joel Lebowitz, and Alex Solon for fruitful discussions. We thank Jessica Metzger and Julia Yeomans for discussions on the many-body implications of our results.
AA acknowledges the financial support of the MathWorks fellowship. AA and JT acknowledge the hospitality of MSC laboratory and the financial support of an MIT MISTI GSF grant.

\appendix
\section{Generalized Langevin equation in $d$ dimensions and for nonzero tracer temperature} \label{appen:ddimgenlan}

Here, we generalize the results of Sec.~\ref{sec:genlan} to $d$ dimensions, and furthermore allow for a nonzero tracer temperature $T_0>0$.
We let $\br \in \mathbb{R}^d$ denote the tracer position. As before, we pack all bath degrees of freedom into a vector $\x \in \mathbb{R}^{M}$; if there are $N$ bath particles in $d$ dimensions, then $M = dN$. The most general linear coupling can be expressed in the form
\begin{equation}\label{eq:ddimgenlanham}
\mathcal{H}(\br, \mathbf{x}) = \frac{1}{2} \mathbf{x}^T A \mathbf{x} + \br^T C \mathbf{x} + \tilde{U}(\br),\end{equation}
where $A$ and $C$ are $M \times M$ and $d \times M$ matrices, respectively. The equations of motion are then
\begin{align}
    \dot{\br} &= -\mu \nabla \tilde{U} - \mu C \mathbf{x} +\sqrt{2 \mu T_0} \bm{\zeta}(t), \label{eq:tracerGeneralLinearddim}
    \\
    \dot{\x} &= -A \x - C^T \br +\sqrt{2T} \bm{\eta}(t),
\end{align}
where $\bm{\eta} \in \mathbb{R}^M$ and $\bm{\zeta} \in \mathbb{R}^d$ are independent centered Gaussian white noises with correlations $\langle \eta_i(t) \eta_j(t') \rangle = \langle \zeta_i(t) \zeta_j(t') \rangle = \delta_{ij}\delta(t-t')$. As in the one-dimensional case, we integrate the bath dynamics to obtain a closed equation for the tracer
\begin{equation}\label{eq:qdot}
\dot{\br} = -\mu \nabla \tilde{U} + \mu \int_{-\infty}^t ds \, C e^{-A(t-s)} C^T \br(s) + \mu \bm{\xi}(t),\end{equation}
where $\bm{\xi}(t) \equiv \sqrt{2 \mu T_0} \bm{\zeta}(t) + \sqrt{2T}\int^t ds \, C e^{A(t-s)} \bm{\eta}(s)$. This can be recast into a generalized Langevin equation using the same manipulations as in Sec.~\ref{sec:genlan}. The result is:
\begin{equation} \label{eq:ddimGenLan}
    \int_{-\infty}^t ds K(t-s) \dot{\br}(s) = -\nabla \tilde{U}(\br(t))+\bm{\xi}(t),
\end{equation}
where
\begin{align}
    K(s) &\equiv 2\mu^{-1} \delta(s) \mathbb{1} + C A^{-1}\left(e^{-A|s|}-1\right) C^T, \\
    G(s) &\equiv \langle\bm{\xi}(t+s) \otimes \bm{\xi}(t)\rangle \nonumber \\
    &= 2\mu^{-1} T_0 \delta(s) \mathbb{1} + T C A^{-1} e^{-A|s|} C^T.
\end{align}
When $T_0 = T$, we have an FDT upon shifting the potential as:
\begin{equation} U(\br) = \tilde{U}(\br) - \frac{1}{2}\br^T C A^{-1} C^T \br.
\end{equation}
It can then be verified that the analysis of the fully-connected model in Sec.~\ref{subsec:genLanMF} generalizes without modification to higher dimensions: the friction and noise kernels are isotropic and given by Eqs.~\eqref{eq:unrescaledkernels}. If the tracer temperature is nonzero, this simply contributes a term $2\mu^{-1}T_0 \delta(s)$ to Eq.~\eqref{eq:connectedNoiseKern}. Since this term is $\mathcal{O}(N^{0})$, the large-$N$ results are unmodified.
\section{Perturbation theory for the fully-connected model}
\subsection{Proof that $H_n$ is a polynomial of degree $n+1$} \label{appen:polynomialHn}
Here, we prove that the expansion coefficients $\{H_n\}$ of the effective Hamiltonian in Eq.~\eqref{eq:effectiveHamExpansion} are polynomials in $p$ of degree $n+1$, so that $h^{(n)}_m = 0$ for $m>n+1$. The proof is by induction in $n$ with Eq.~\eqref{eq:H1sol} as the $n=1$ base case. Suppose the statement is true for all $H_2, \dots, H_{n-1}$. Substituting $m = n+2+k$ into Eq.~\eqref{eq:coefficientRecursion} with $k$ a nonnegative integer, we obtain
\begin{multline} T(n+3+k)(n+4+k) h_{n+4+k}^{(n)}-(n+2+k) h_{n+2+k}^{(n)} 
\\= \sum_{\ell=0}^{n+2+k} (n+k-\ell+3)(\ell+1) \sum_{q=1}^{n-1} h_{n+k-\ell+3}^{(q)} h_{\ell+1}^{(n-q)}\;.
\label{eq:terminationAnsatz}
\end{multline}
Assume, for the sake of contradiction, that there exists nonzero terms in the sum on the RHS of Eq.~\eqref{eq:terminationAnsatz}. By the induction hypothesis, such terms must satisfy both $n+k-\ell +2 \leq q$ and $\ell \leq n-q$. Adding these inequalities leads to a contradiction, implying that the RHS of Eq.~\eqref{eq:terminationAnsatz} is zero, and hence that
\begin{equation} h^{(n)}_{n+4+k} = \frac{n+2+k}{T(n+3+k)(n+4+k)}h^{(n)}_{n+2+k}. \label{eq:hnkrecursion}\end{equation}
We now partition the power series for $H_n$ as
\begin{align} H_{n} &= \sum_{m=0}^{n+1} h^{(n)}_m p^m + p^{n+2} G_n(x,p),\\
G_n(x, p) &\equiv \sum_{k=0}^\infty h^{(n)}_{n+2+k}(x) p^k.
\end{align}
Using Eq.~\eqref{eq:hnkrecursion}, we rewrite $G_n(x,p)$ as
\begin{equation} \label{eq:gdef}
    G_n(x,p) = h_{n+2}^{(n)} \sum_{\ell=0}^\infty a_\ell \left(\frac{p^2}{2T}\right)^\ell + p h_{n+3}^{(n)} \sum_{\ell=0}^\infty b_\ell \left(\frac{p^2}{2T}\right)^\ell,
\end{equation}
where we have separated the sum into its even and odd parts, and where $\{a_\ell, b_\ell\}$ are defined recursively by $a_0 = b_0 = 1$ and
\begin{align}\frac{a_{\ell+1}}{a_\ell} &= \frac{\left(\frac{n+2}{2}+\ell\right)}{\left(\frac{n+3}{2}+\ell\right)\left(\frac{n+4}{2}+\ell\right)},\\ \frac{b_{\ell+1}}{b_\ell} &= \frac{\left(\frac{n+3}{2}+\ell\right)}{\left(\frac{n+4}{2}+\ell\right)\left(\frac{n+5}{2}+\ell\right)}\;.
\end{align}
Since these are rational functions of $\ell$, we identify the sums in~\eqref{eq:gdef} as generalized hypergeometric series \cite{arfken2011mathematical}
\begin{multline} \label{eq:hyperseries}
    G_n(x,p) = h_{n+2}^{(n)} \, {}_2 F_2\left[\frac{n+2}{2}, 1 ; \frac{n+3}{2}, \frac{n+4}{2}; \frac{p^2}{2T}\right] \\+ p h_{n+3}^{(n)} \, {}_2 F_2\left[\frac{n+3}{2}, 1 ; \frac{n+4}{2}, \frac{n+5}{2}; \frac{p^2}{2T}\right].
\end{multline}
The leading terms in a large-$p$ expansion of the above are
\begin{multline} \label{eq:GnLarge}
    G_n(x,p) =  e^{\frac{p^2}{2T}} \frac{T^{\frac{n+3}{2}}}{p^{n+3}} \sqrt{\frac{\pi}{2}} \Big[(n+2)(n+1)!! \, h^{(n)}_{n+2} \\+ \sqrt{T}(n+3)(n+2)!! \, h^{(n)}_{n+3} + \cO(p^{-1})\Big].
\end{multline}
We see that $G_n$ (and hence $H_n$) grows exponentially in $p^2$ for large $p$, so that the distribution $\Psi(x,p) \propto e^{-\mathcal{H}/T}$ is not normalizable in $p$ unless $G_n = 0$. This implies that $h_{n+2}^{(n)} = h_{n+3}^{(n)} = 0$, so that by Eq.~\eqref{eq:hnkrecursion}, we have
\begin{equation} \label{eq:hpoly}
h^{(n)}_{m} = 0 \quad \text{for all} \quad m > n+1. \end{equation} 
We conclude that $H_n$ is a polynomial in $p$ of degree $n+1$.

\subsection{Expectation values in perturbation theory} \label{appen:expec}
We show here how expectation values can be computed perturbatively in powers of $\eps$. Using $H_0 = p^2/2+U(x)$, we may write the probability density as 
\begin{align} \Psi(x,p) &= \frac{1}{Z} \exp{\left[-\frac{U(x)}{T} - \frac{p^2}{2T}-\frac{1}{T}\sum_{n=1}^\infty \eps^n H_n(x,p)\right]} \nonumber\\
&= \frac{e^{-U/T-p^2/2T}}{Z_0}\left[1+\sum_{n=1}^\infty \eps^n A_n(x,p)\right],
\label{eq:exponentialExpansion}\end{align} 
where $Z_0 \equiv \int dx dp \, e^{-H_0(x,p)/T} = \sqrt{2 \pi T} \int dx \, e^{-U(x)/T}$ is the equilibrium partition function, and where we have imposed the normalization~\footnote{Note that Eq.~\eqref{eq:normalizationAn} is another way to show that $G_n$ in Eq.~\eqref{eq:GnLarge} must vanish.}
\begin{equation} 
\int dx dp\, e^{-U/T-p^2/2T} A_n(x,p) = 0. \label{eq:normalizationAn} 
\end{equation}
Each $A_n$ is obtained from $\{H_n, H_{n-1}, \dots, H_1\}$ by expanding the exponential and matching powers of $\eps$. So long as the integral on the left hand side of~\eqref{eq:normalizationAn} is finite, the normalization condition~\eqref{eq:normalizationAn} on $A_n$ can be satisfied by adding constants to the $\{H_n\}$, which does not change $\Psi(x,p)$. The choice~\eqref{eq:normalizationAn} keeps $\Psi$ exactly normalized at each order in $\eps$, so that that we need only compute the equilibrium partition function $Z_0$. Since $H_n$ is a polynomial in $p$ of degree $n+1$, $A_n$ will be a polynomial in $p$ of degree $2n$ due to the contribution proportional to $(H_1)^{n}$:
\begin{equation}\label{eq:Anofan}
A_n = \sum_{m=0}^{2n} p^m a^{(n)}_m(x)
\end{equation}
Upon computing the $\{A_n\}$, any expectation value can be expressed as
\begin{equation} \label{eq:expecVals}
\langle f(x,p) \rangle = \langle f(x,p) \rangle_0 + \sum_{n=1}^\infty \eps^n \langle f(x,p) A_n(x,p) \rangle_0,
\end{equation}
where $\langle \cdot \rangle_0$ denotes an expectation with respect to the equilibrium $(\eps=0)$ measure with density $e^{-U/T-p^2/2T}/Z_0$. In particular, we will later be interested in computing moments of the momentum variable $p$. The formula for Gaussian central moments tells us
\begin{equation}\langle p^k \rangle_0 = \begin{cases}
    0, & \text{for } k \text{ odd}, \\
    T^{k/2} (k-1)!!, & \text{for } k \text{ even}.
\end{cases}\end{equation}
Using Eqs.~\eqref{eq:Anofan} and~\eqref{eq:expecVals}, this leads to
\begin{equation}\langle p^k \rangle = \sum_{n=0}^\infty \eps^n \sum_{m=0}^n T^{m+\frac{k+b}{2}}(k + 2m + b - 1)!! \left\langle a_{2m+b}^{(n)}\right\rangle_0
\label{eq:pmoment}
\end{equation}
where $b = 0$ if $k$ is even and $b=1$ if $k$ is odd, and where $a^{(n)}_{m>2n} = 0, a^{(0)}_0 = 1$.

We would also like to compute the marginal density for the tracer position $P(x') \equiv \langle \delta(x'-x)\rangle$, which can be expressed as
\begin{align} P(x) &= \int_{-\infty}^\infty dp \, \Psi(x,p), \nonumber \\
&= \frac{e^{-U(x)/T}}{Z_{0, x}} \left[1 + \sum_{n=1}^\infty \eps^n \int dp \frac{e^{-p^2/2T}}{\sqrt{2 \pi T}} A_n(x,p)\right], \nonumber \\
&= \frac{e^{-U(x)/T}}{Z_{0, x}} \left[1 + \sum_{n=1}^\infty \eps^n F_n(x)\right], \label{eq:marginalGen}
\end{align}
where $Z_{0,x} \equiv \int dx \, e^{-U(x)/T}$, and
\begin{equation} F_n(x) \equiv \sum_{m=0}^n T^{m}(2m-1)!!\, a_{2m}^{(n)}(x). \label{eq:Fndef}
\end{equation}
\subsection{Calculating the $\{\hnp{n}{m}\}$} \label{appen:hnp}
Using Eq.~\eqref{eq:hpoly} and substituting $m = n+1$ into Eq.~\eqref{eq:coefficientRecursion} leads to
\begin{equation}h^{(n)}_{n+1}(x) = -\frac{1}{\sqrt{k}(n+1)} \frac{d h_{n}^{(n-1)}}{dx}.\end{equation}
This recursion relation can be solved with $h^{(1)}_2 = (k^{-1}U''+1)/2$ as initial condition (Eq.~\ref{eq:H1sol}). The result is
\begin{equation} 
h^{(n)}_{n+1} = \frac{(-1)^{n-1}}{k^{n/2+1/2}(n+1)!}U^{(n+1)}(x), \qquad n \geq 2. \label{eq:hnnp1}
\end{equation}
Next, we substitute $m = n$ into Eq.~\eqref{eq:coefficientRecursion} and use Eq.~\eqref{eq:hnnp1}  to get
\begin{equation}
n h_{n}^{(n)} + k^{-1/2} \partial_x h_{n-1}^{(n-1)} = V_n(x), \qquad n \geq 2, 
\end{equation}
where $V_2 \equiv 0$ and
\begin{equation}V_{n>2}(x) \equiv
\frac{(-1)^{n-1} \!}{k^{n/2}}\!\left[ \frac{U^{(n)}}{(n\!-\!1)!}\! +\! \frac{1}{k} \sum_{\ell=1}^{n-2} \frac{U^{(n-\ell+1)} U^{(\ell+1)}}{\ell! (n-\ell)!} \right],\end{equation}
Using the initial condition $h_1^{(1)} = 0$ from Eq.~\eqref{eq:H1sol}, the solution can be verified to be
\begin{equation} h_n^{(n)} = \sum_{\ell=0}^{n-2} \frac{(n-\ell-1)!}{n! \,k^{\ell/2}}\left(-1\right)^\ell V^{(\ell)}_{n-\ell}(x). \label{eq:hnn}
\end{equation}
We can continue like this, substituting $m = n-k$ and deriving closed recursive systems of ODEs for $h^{(n)}_{n-k}$ in terms of $h^{(n)}_{n-k}, h^{(n)}_{n-k+1}, \dots, h^{(n)}_{n+1}$. However, the recursion relation for $h^{(n)}_{n-1}$ will have no initial condition because $h^{(1)}_0$ remains undetermined. More generally, solving for $h^{(n)}_{n-k}$ requires knowledge of $h^{(k)}_0$. Direct substitution of $m=0$ into Eq.~\eqref{eq:coefficientRecursion} does not yield an equation for $h^{(n)}_0$ due to the factors of $m$ which multiply $h^{(n)}_m$ and $h^{(n-1)}_m$. Although it is not possible to write a general expression for $h^{(n)}_0$, one can systematically find $h^{(n)}_0$ for successively higher values of $n$ by solving Eq.~\eqref{eq:coefficientRecursion} for $h^{(n+1)}_{m>0}$ and $h^{(n+2)}_{m>1}$. We now demonstrate this procedure for $n=2$ and $n=3$. 

Starting with $n=2$, equations~\eqref{eq:hnnp1} and~\eqref{eq:hnn} give
\begin{equation}\hnp{2}{3} = -\frac{1}{6k^{3/2}}U^{(3)}(x), \qquad \hnp{2}{2} = 0.\end{equation}
Equation~\eqref{eq:coefficientRecursion} with $m=1$ then yields
\begin{equation} \partial_x \hnp{1}{0}=- \sqrt{k} \hnp{2}{1}-T k^{-1} U^{(3)} + U'(k^{-1} U'' + 1). \label{eq:h10ode} \end{equation}
(Note that we deviate from our convention for clarity and use $\partial_x$ to refer to $\hnp{1}{0}{}'(x)$ since $\hnp{n}{p}$ has an upper index.) For $n=3$, we obtain from Eqs.~\eqref{eq:hnnp1} and~\eqref{eq:hnn}
\begin{equation}\hnp{3}{4} = \frac{1}{24 k^2}U^{(4)}, \qquad \hnp{3}{3} = \frac{1}{6 k^{3/2}}U^{(3)}\left(1+k^{-1} U^{(2)}\right).\end{equation}
Setting $m=2$, $m=1$ and $m=0$ in Eq.~\eqref{eq:coefficientRecursion} then yields, respectively
\begin{align}
\partial_x \hnp{2}{1} &= -2\sqrt{k} \hnp{3}{2} + \frac{T}{2 k^{3/2}} U^{(4)} - \frac{1}{2 k^{3/2}} U' U^{(3)}, \\
\partial_x \hnp{2}{0} &= \frac{T}{k} U^{(3)} \left(1+\frac{U''}{k}\right) - \sqrt{k} \hnp{3}{1} - \left(1+\frac{U''}{k}\right) \hnp{2}{1},  \label{eq:h20ode}\\
\hnp{3}{2} &= - \frac{1}{2T\sqrt{k}} U' \hnp{2}{1}.
\end{align}
Substituting the last of these into the first gives a closed equation for $\hnp{2}{1}$:
\begin{equation}\partial_x \hnp{2}{1} = \frac{U'}{T} \hnp{2}{1} + \frac{T}{2 k^{3/2}} U^{(4)} - \frac{1}{2 k^{3/2}} U' U^{(3)}.\end{equation}
The solution is
\begin{equation}\hnp{2}{1} = c_1\,e^{U/T} + \frac{T}{2 k^{3/2}} U^{(3)},\end{equation}
where $c_1$ is an integration constant whose value will be determined shortly. Plugging into~\eqref{eq:h10ode}, we find
\begin{equation}
\hnp{1}{0} = -  \frac{3T}{2k} U'' + \frac{(U')^2}{2k} + U -\sqrt{k}c_1 \int^x dx \, e^{U/T}.\end{equation}

In an infinite system with a confining potential, the last term above leads to a violation of the normalization condition~\eqref{eq:normalizationAn} for $n=1$. In a finite system, this term is forbidden by periodic or reflecting boundary conditions. We conclude that $c_1 = 0$, and hence
\begin{align}\hnp{2}{1} &= \frac{T}{2k^{3/2}}U^{(3)}, \qquad \hnp{3}{2} = -\frac{1}{4k^2} U' U^{(3)},
\nonumber\\\hnp{1}{0} &= - \frac{3T}{2k} U'' + \frac{(U')^2}{2k} + U.\end{align}
We now proceed to $n=4$. Equations~\eqref{eq:hnnp1} and~\eqref{eq:hnn} give
\begin{align}
\hnp{4}{5} &= -\frac{U^{(5)}}{120k^{5/2}}, \\
\hnp{4}{4} &= -\frac{1}{48k^2}\left[\frac{\big(U^{(3)}\big)^2}{k} + 4U^{(4)}\left(1+ \frac{U''}{k}\right)\right].
\end{align}
Substituting $n=4$ and $m=3$ into~\eqref{eq:coefficientRecursion} gives:
\begin{multline}
    \hnp{4}{3} =-\frac{U^{(3)} }{4 k^{7/2}} \left[k U''+\frac{2}{3} \left(k^2 + U''^2\right)\right]\\
    +\frac{5 U^{(4)} U'-2 T U^{(5)}}{36 k^{5/2}}
\end{multline}
The $m=2$ and $m=0$ equations can together be solved to give:
\begin{align*}
    \hnp{4}{2} &= \frac{T (U^{(3)})^2}{8 k^3}-\frac{U'}{2 \sqrt{k} T} \hnp{3}{1}  \\
    \hnp{3}{1} &= -\frac{T U^{(3)}}{k^{3/2}}   \left(1+ \frac{U''}{k}\right)+c_2 e^{\frac{U}{T}},
\end{align*}
This can be substituted into~\eqref{eq:h20ode}, and by the same argument we used to set $c_1=0$, we find $c_2=0$. We conclude:
\begin{align}
     \hnp{3}{1} &= -\frac{T U^{(3)} }{k^{3/2}}  \left(1+ \frac{U''}{k}\right), \\
     \hnp{4}{2} &= \frac{U^{(3)}}{8 k^3} \left(4 U' \left(k+U''\right)+T U^{(3)}\right), \\
     \hnp{2}{0} &= \frac{3 T}{4 k^2} U'' \left(2 k+U''\right).
\end{align}
Finally, we will look at $n=5$. From~\eqref{eq:hnnp1} and~\eqref{eq:hnn},
\begin{align}
\hnp{5}{6} &= \frac{U^{(6)}}{720 k^3}, \\
\hnp{5}{5} &= \frac{1}{120 k^{7/2}} \left(3 U^{(5)} \left(k+U''\right)+11 U^{(3)} U^{(4)}\right).
\end{align}
Equation~\ref{eq:coefficientRecursion} with $n = 5$ and $m=4,3$ can be used to determine $\hnp{5}{4}$ and $\hnp{5}{3}$, using which  the $m=2$ ODE can be solved together with $m=0$ to find $\hnp{4}{1}$ and $\hnp{5}{2}$. The result for $\hnp{4}{1}$ is then substituted into the $n=4$, $m=1$ equation and the ODE can be solved for $\hnp{3}{0}$. The results of these calculations are
\begin{widetext}
\begin{align}
    \hnp{5}{4} =& \frac{1}{288 k^4}\!\left\{k \left(7 T U^{(6)} \! -\! 13 U^{(5)} U'\right)+4 U^{(4)} \left[9 k^2 \!+ \! 14 k U''\!+\!9 U''^2\right]
    +\big(U^{(3)}\big)^2 \!\left[90 U''+84 k \right]\right\}.  \\
    \hnp{5}{3} =& \frac{\left(k+U''\right)}{36 k^{7/2}}  \left(8 T U^{(5)}-15 U^{(4)} U'\right)
    +\frac{U^{(3)} \left(k+U''\right)}{12 k^{9/2}}   \left[2 k^2+k  U''+2 U''^2\right]
    +\frac{17}{36 k^{7/2}} \left[T U^{(3)} U^{(4)} - \big(U^{(3)}\big)^2 U'\right] \\
    \hnp{4}{1} =& \, c_3 e^{U/T} + \frac{3  T U^{(3)} U''^2}{2 k^{7/2}}+\frac{11T U^{(3)} U''}{4k^{5/2}} +\frac{7 T^2 U^{(5)}}{24 k^{5/2}}+\frac{3 T U^{(3)}}{2 k^{3/2}}-\frac{5 T U^{(4)} U'}{12 k^{5/2}}, \\
    \hnp{5}{2} =& -\frac{U' }{2 \sqrt{k} T} \hnp{4}{1} -\frac{T (U^{(3)})^2 U''}{2 k^4}-\frac{T (U^{(3)})^2}{2 k^3}, \\
\hnp{3}{0}=& -\frac{T}{2k^3}\left(1+\frac{7k}{2}\right)U''^3-\frac{3}{2}\frac{T}{k}U'' -\frac{5 T \left(T U^{(4)}-2 U^{(3)} U'\right)}{8 k^2} -\frac{1}{2 k^2} \int^x dz\, U^{(3)}(z) U'^2(z)-\sqrt{k} c_3 \int^x dz\, e^{U(z)/T}
\end{align}
\end{widetext}
In an infinite domain with a confining potential, normalization requires $c_3 = 0$ for the reasons discussed earlier. This is no longer true in a finite domain $(0,L)$ with periodic boundaries: using $U^{(n)}(0) = U^{(n)}(L)$, we have
\begin{align}0 &= \hnp{3}{0}(L) - \hnp{3}{0}(0) \nonumber\\
&=-\frac{1}{2 k^2} \int_0^L dz\, U^{(3)}(z) U'^2(z)-\sqrt{k} c_3\int_0^L dz\, e^{\frac{U(z)}{T}},
\end{align}
so that for periodic boundaries
\begin{equation}c_3 = -\frac{1}{2k^{5/2}} \frac{\int_0^L dz\, U^{(3)}(z) U'^2(z)}{\int_0^L dz\, e^{U(z)/T}}.\end{equation}
\subsection{Marginal distribution for the tracer position}
\label{appen:marginal}
Having determined $H_1, H_2$ and $H_3$, we can now calculate $A_1, A_2$ and $A_3$, as defined in Eq.~\eqref{eq:exponentialExpansion}. Since we have solved for $\hnp{4}{m>0}$, we can also compute the momentum-dependent parts of $A_4$. In order to satisfy the normalization condition~\eqref{eq:normalizationAn}, we must add to each $H_n$ a constant 
\begin{equation}H_n \rightarrow H_n + b_n,\end{equation}
with $b_n$ determined by~\eqref{eq:normalizationAn}. This modifies only the normalization prefactor of $\Psi$. Expanding~\eqref{eq:exponentialExpansion} in $\eps$, matching powers of $\eps$, and then matching powers of $p$, we can express $\anp{n}{m}$ in terms of the $\{\hnp{k}{p}\}$ with $k \leq n$. 
The marginal distribution for the tracer position can then be computed from~\eqref{eq:marginalGen} using the coefficients $F_n$ defined in Eq.~\eqref{eq:Fndef}. The first two are given by
\begin{subequations}\begin{align}
     F_1 =& -\frac{1}{2}-\frac{b_1}{T}-\frac{U}{T}-\frac{U'^2}{2kT}+\frac{U''}{k}, \\
    F_2 =& \frac{b_1(T+b_1)-2Tb_2}{2T^2} -\frac{U''}{2 k T}\left(2 b_1 +3 T+2 U+\frac{U'^2}{k}\right) \nonumber \\ &+ \frac{\left( U+\frac{U'^2}{2k}\right) \left(2 b_1+T+U+\frac{U'^2}{2k}\right)}{4 k T^2} + \frac{3}{8}
\end{align}\end{subequations}
The expression for $F_3$ is omitted for brevity. We can re-exponentiate the series to express this in terms of an effective potential $P(x) \propto e^{-U_\mathrm{eff}/T}$:
\begin{align}
P(x) &= \frac{e^{-U(x)/T}}{Z_{0, x}} \left[1 + \sum_{n=1}^\infty \eps^n F_n(x)\right] \nonumber\\
&\propto \exp\left[-\frac{U(x)}{T}-\frac{1}{T}\sum_{n=1}^\infty \eps^n U_n(x)\right],
\end{align}
Matching orders of $\eps$ yields Eqs.~\eqref{eq:marginalEffectivePot}, where all $x$-independent constants have been discarded from $U_n$ because they can be absorbed into the normalization prefactor. 



\subsection{Generalization to $d$ dimensions}\label{appen:ddimperturb}
Several results from the perturbation theory in Sec.~\ref{sec:active} may be generalized to $d$ spatial dimensions. The Fokker-Planck equation becomes $\partial_t \Psi(\br, \bp) = \mathcal{L} \psi(\br, \bp)$, with 
\begin{align} \label{eq:fokkerplanckddim}
    \eps \mathcal{L} =& \eps \sqrt{k} \left(\nabla U \cdot \nabla_\bp - \bp \cdot \nabla\right) \nonumber \\&+ \nabla_\bp \cdot \left[\eps \nabla (\bp\cdot\nabla)U + k(1+\eps)\bp\right] + kT \nabla_\bp^2.
\end{align}

We first note that if the potential is one-dimensional, so that $\nabla U = U'(x) \hat{\x}$ along some direction $\hat{\x}$, then Eq.~\eqref{eq:fokkerplanckddim} reduces to Eq.~\eqref{eq:fokkerplanck} and the problem becomes effectively one-dimensional. Our results on the ratchet current (Eq.~\ref{eq:current}) and density rectification (Eq.\ref{eq:rectification}) thus generalize trivially to $d$ dimensions, so long as the asymmetric external potential is unidirectional. 

In the general case, more work is required to compute the correction to the Boltzmann distribution and the entropy production rate. The $d$-dimensional generalization of Eq.~\eqref{eq:hammypde} is
\begin{align}
    \left\{\eps \sqrt{k} \left(\nabla U \!\cdot\! \nabla_\bp - \bp \!\cdot \!\nabla\right) + \left[\eps (\bp \!\cdot \!\nabla) \nabla U + k(1\!+\!\eps)\bp\right]\cdot \nabla_\bp \right. \nonumber\\
    + kT \nabla_\bp^2\Big\} \mathcal{H}(\br, \bp) = k\left(\nabla_\bp \mathcal{H} \right)^2 +\eps T \nabla^2 U + Td k(1+\eps) 
\end{align}
As before, we expand $\mathcal{H} = \sum_{n=0}^\infty \eps^n H_n(\br, \bp)$, and obtain the following recursion relation for $n \geq 2$:
\begin{align}
    \nonumber \left[T\nabla_{\bp}^2 - \bp \cdot \nabla_\bp\right] H_n + \bigg\{ \!\left[\bp+  \left(k^{-1/2}+\frac{\bp\cdot\nabla}{k}\right)\nabla U\right]\!\cdot\! \nabla_\bp \\
    - k^{-1/2}\bp \cdot \nabla \bigg\} H_{n-1} = \sum_{m=1}^{n-1}\left(\nabla_\bp H_m\right)\left(\nabla_\bp H_{n-m}\right)
\end{align}
The polynomial expansion of $H_n$ in Eq.~\ref{eq:polyExpansion} now has tensorial coefficients:
\begin{equation}
    H_n(\br, \bp) = \sum_k h^{(n,k)}_{\alpha_1 \dots \alpha_k}(\br) \,\, p_{\alpha_1}\cdots p_{\alpha_k}.
\end{equation}
The corresponding generalization of Eq.\eqref{eq:coefficientRecursion} is cumbersome. We instead work out the first few orders by brute-force substitution. The results are listed below, setting $k=1$ for brevity:
\begin{align}
    H_0 =& \frac{p^2}{2} + U, \\
    H_1 =& U + \frac{\left(\nabla U\right)^2}{2} - \frac{3}{2}T \nabla^2 U + \frac{1}{2}\left(\bp \cdot \nabla\right)^2 U + \frac{p^2}{2}, \\
    H_2 =& f_2(\br) + \frac{T}{2}(\bp\cdot\nabla)\nabla^2 U - \frac{1}{6} (\bp \cdot \nabla)^3 U, \\
    H_3 =& f_3(\br) - \bp \cdot \nabla f_2 - \frac{T}{2} (\bp \cdot \nabla) \nabla^2 U  + \frac{1}{6}(\bp \cdot \nabla)^3 U, \nonumber\\
    &- \frac{1}{4} (\nabla U)\cdot (\bp \cdot \nabla)^2 \nabla U +\frac{1}{24}(\bp\cdot\nabla)^4 U  \nonumber\\
    &+ T p_\alpha\! \left[(\partial_\beta \partial_\lambda U)(\partial_\beta \partial_\alpha \partial_\lambda U) \!-\! \frac{3}{2}(\partial_\alpha \partial_\beta U)(\partial_\beta \nabla^2 U)\right] \nonumber \\
    &+ \frac{1}{6} p_\alpha p_\beta p_\gamma (\partial_\alpha \partial_\lambda U)(\partial_\beta \partial_\gamma \partial_\lambda U).
\end{align}
with $f_2$ and $f_3$ undetermined at this order. Using these results, we compute the marginal distribution of the tracer position as $p(\br)\propto e^{-U_\mathrm{eff}(\br)/T}$, where
\begin{equation}
    U_\mathrm{eff}(\br) = U+ \eps\left[\frac{\left(\nabla U \right)^2}{2} - T \nabla^2 U + U\right] + \mathcal{O}(\eps^2).
\end{equation}

Finally, we generalize the result on the entropy production rate. The dynamical action of Eq.~\eqref{eq:action} reads
\begin{equation}
    S = \int^t_0 ds \frac{\left[\dot{\bp} + \nabla U + (\bp \cdot \nabla) \nabla U + (N+1)\bp \right]}{4TN}
\end{equation}
Using similar manipulations as in Eqs. \eqref{eq:SRminusS}-\eqref{eq:entropy}, we obtain
\begin{equation}
    \sigma = \frac{\eps}{2T} \left \langle \left(\bp \cdot \nabla \right)^3 U \right\rangle.
\end{equation}
To compute this, we expand as before
\begin{equation}
    \sigma = \frac{\eps}{2T} \left \langle \left(\bp \cdot \nabla \right)^3 U \left[1+\eps A_1 + \eps^2 A_2 \right] \right\rangle_0 + \mathcal{O}(\eps^4),
\end{equation}
where $A_1(\br, \bp) = H_1/T$ and $A_2(\br, \bp) = H_1^2/2T^2-H_2/T$. Upon evaluating the Gaussian averages over $\bp$, we obtain
\begin{equation}
    \sigma = \frac{\eps^3 T}{2} \left\langle \left(\nabla^2 U \right)^2 \right\rangle + \mathcal{O}(\eps^4).
\end{equation}
This scales as $N^{-3}$, proving that the time-reversible yet non-Boltzmann regime extends to higher dimensions.

\section{Weakly non-linear external potentials} \label{appen:nonlinear}
Here, we show how the general formalism of Sec.~\ref{sec:exact}, and in particular Eqs.~\eqref{eq:momentdev} and~\eqref{eq:correlRat}, can be extended to weakly anharmonic potentials acting on the tracer. We thus consider a potential energy:
\begin{equation} \label{eq:potential}
\cH(\x) = \frac{1}{2}\x^T A \x + \eps Q(\x),
\end{equation}
where $\eps$ is taken to be small. The equations of motion~\eqref{eq:linearEOM} become $\dot{\x} = -Z \x - M \nabla H + \mathbf{\Gamma}$, so their solution read
\begin{equation}\x = \x\bare - \eps \int_{0}^t ds G(t-s) M \nabla Q(\x(s)),\end{equation}
where $\x\bare$ is a solution to the linear theory with $\eps = 0$, and $G$ is the unperturbed Green's function given in Eq.~\eqref{eq:greendef}. To leading order in $\eps$, the value of any two-point correlator is then given by
\begin{align}
\langle x_i x_j \rangle \!=\! \langle x_i x_k \rangle_0 &\!-\! \eps \!\int_0^t \!ds \mu_k\left[G_{jk}(t\!-\!s)\langle x_i(t) \partial_k Q(\x(s)) \rangle _0 \right. \nonumber
\\ &\left. - G_{ik}(t\!-\!s) \langle x_j(t) \partial_k Q(\x(s))\rangle_0 \right], \label{eq:GeneralTwopointNonlinear}
\end{align}
where $\langle \cdot \rangle_0$ denotes an expectation value with respect to the Gaussian measure of the $\eps = 0$ linear theory. For simplicity, we focus on the case where the nonlinearity acts only on the tracer. In the notation of Eqs.~\eqref{eq:genEOM}, this means that $V$ is quadratic but $U$ is not: $\eps Q(\mathbf{x}) = U(x_0)$. We pick a quartic form for $U$:
\begin{equation}U(x_0) = \frac{\eps}{4}x_0^4.\end{equation}
It then follows from Eq.~\eqref{eq:GeneralTwopointNonlinear} and Wick's theorem that
\begin{align}
\langle x_i x_j \rangle = \langle x_i x_k \rangle_0 - 3\eps \!\int_0^t \!ds \mu_0\left[G_{j0}(t\!-\!s)\langle x_i(t) x_0(s) \rangle _0 \right. \nonumber
\\ \left. - G_{i0}(t\!-\!s) \langle x_j(t) x_0(s)\rangle_0 \right]\left\langle x_0(s)^2\right\rangle, \label{eq:QuarticTwopointNonlinear}
\end{align}
To proceed, we note that in the linear theory, unequal-time correlators are related to equal-time correlators through the (noiseless) Green's function \cite{risken1996}. In particular, defining the matrix $K_{ij}(s, t) = \langle x_i(t+s) x_j(t) \rangle_0$, it is straightforward to verify that $K(s, t) = G(s) K(0, t)$. Using this fact, Eq.~\eqref{eq:QuarticTwopointNonlinear} can be shown to imply the following form for the steady-state two-point correlator,
\begin{equation} \label{eqn:twopoint-greens}
\langle x_i x_j \rangle = \langle x_i x_j \rangle_0 - 3\epsilon \langle x_0^2 \rangle_0 \langle x_k x_0\rangle_0 B_{ijk}+ \mathcal{O}(\eps^2),\end{equation}
where
\begin{equation}B_{ijk} \equiv \mu_0 \int_0^\infty ds \left[G_{j0}(s) G_{ik}(s)+G_{i0}(s)G_{jk}(s)\right].\end{equation}
As done in section~\ref{sec:exact}, we may diagonalize $G$ and express $B_{ijk}$ explicitly in terms of the eigenvalues and eigenvectors of $Z$. We find
\begin{equation}B_{ijk} = \mu_0 \sum_{n,m=0}^N \frac{u_0^{(n)}u_k^{(m)}}{\lambda_n + \lambda_m}\left[v_{j}^{(n)}  v_{i}^{(m)} + v_{i}^{(n)}  v_{j}^{(m)}\right].\end{equation}
where we remind that $\{\mathbf{v}^n\}$ and $\{\mathbf{u}^n\}$ are respectively the right and left eigenvectors of $Z$. 

Upon computing Eq.~\eqref{eqn:twopoint-greens} for a particular model, we would like to compare the result to what it would be in an equilibrium ensemble. The equilibrium expectation value of any observable $f(\x)$ can be computed perturbatively in $\eps$ following standard procedures
\begin{align}
\langle f(\x) \rangle^\eq &=\frac{\int d\x f(\x) \exp\left[-\frac{1}{2T}\x^T A\x - T^{-1} U(x_0)\right]}{\int d\x\exp\left[-\frac{1}{2T}\x^T A\x - T^{-1} U(x_0)\right]} \nonumber \\
&= \langle f\rangle_0^\eq - T^{-1}\left(\langle f U\rangle_0^\eq - \langle f\rangle_0^\eq \langle U\rangle_0^\eq\right) +\mathcal{O}(\eps^2),
\end{align}
using which it can be shown that,
\begin{equation}\langle x_i x_j \rangle^{\mathrm{eq}} = T A^{-1}_{ij} - 3 \epsilon T^2 A^{-1}_{i0} A^{-1}_{j0} A^{-1}_{00} + \mathcal{O}(\epsilon^2).\end{equation}
This allows us to write an explicit expression for the ratio of equilibrium and nonequilibrium correlators, as was done in the linear case:
\begin{align}R_{ij} \equiv \frac{\langle x_i x_j \rangle}{\langle x_i x_j \rangle^\eq} = R_{ij} \bare \left\{1 + \eps \left[\frac{A^{-1}_{i0} A^{-1}_{j0} A^{-1}_{00}}{A^{-1}_{00}}\right. \right.\nonumber\\
\left. \left. - \frac{\big(A^{-1}_{00}-\Delta\bare_{00}\big)\big(A^{-1}_{k0}-\Delta\bare_{k0}\big)B_{ijk}}{A^{-1}_{ij} - \Delta_{ij} \bare}\right] \right\}, \label{eq:RijPerturb}\end{align}
where $R\bare$ is the $\eps = 0$ value of $R$ given by Eq.~\eqref{eq:correlRat} and $\Delta\bare$ is defined by Eq.~\eqref{eq:momentdev}.

As an application of this perturbation theory, it can be shown that for the fully-connected models of section~\ref{subsec:linearmf}, in the cases where $\lim_{N\rightarrow \infty} R_{ij}\bare = 1$, we also have $\lim_{N\rightarrow \infty} R_{ij} = 1 + \cO(\eps^2)$ according to Eq.~\eqref{eq:RijPerturb}. This not only confirms the conclusion of Section~\ref{sec:genlan} that the tracer equilibrates in arbitrary potentials when $N \rightarrow \infty$, but also allows the finite-$N$ corrections to be computed exactly to leading order in the nonlinearity.

\if{
\section{Average energy of the loop model} \label{appen:virial}
In this Appendix, we show how the identity $\sum_{m\in\mathbb{Z}} \Delta T(m)=T$ in the loop model can be obtained without the full exact solution of Sec.~\ref{sec:loopmodelexact}. We first rewrite the left hand side in terms of bond variables $u_{m} \equiv x_{m+1}-x_m$:
\[\sum_{m\in\mathbb{Z}} \Delta T(m) = \sum_{m\in \mathbb{Z}} \left[T - \frac{1}{2} \left \langle u_m^2\right \rangle\right]\]
Using Ito's lemma, we express the dynamics of $\langle u_m^2 \rangle$ as
\[\partial \langle u_m^2 \rangle \]
}\fi

\section{Two-point correlator for an overdamped elastic medium with a cold inclusion} \label{appen:elastic}
We list here the correlator $\langle u_i(\q,t) u_j(\q',t)\rangle$ computed from Eqs.~\eqref{eq:elasticSol} and~\eqref{eq:lambdacor}. First, Eq.~\eqref{eq:lambdacor} directly leads to
\begin{align}
    \left \langle \Lambda^\parallel_i(\q,s) \Lambda^\parallel_j(\q',s') \right \rangle &=\frac{q_i q_j' \q \cdot \q'}{q^2 q'^2} g(\q, \q', s, s')\\
     \left \langle \Lambda^\parallel_i(\q,s) \Lambda_j(\q',s') \right \rangle &= \frac{q_i q_j}{q^2} g(\q, \q', s, s')
\end{align}
where
\begin{equation}g(\q, \q', s, s') \equiv 2T \delta (s - s') \left[(2\pi)^d \delta(\q + \q') - L^{-d}\right]\end{equation}
Using this, long but straightforward algebra leads to:
\begin{align} \label{eq:uiujElastic}
\langle u_i(\q, t) u_j(\q',t) \rangle = T(2\pi)^d \delta(\q+&\q') \bigg[\frac{\delta_{ij}}{q^2} 
- \frac{\kappa}{1+\kappa} \frac{q_i q_j}{q^4}\bigg] \nonumber\\&- 2T L^{-d} H_{ij},
\end{align}
where the influence of the cold tracer is contained within the tensor $H$, defined as
\begin{align}
H_{ij} \equiv &\frac{\delta_{ij}}{q^2 + q'^2} + \frac{q_i q_j}{q^2}\left[\frac{1}{(1+\kappa) q^2+q'^2}-\frac{1}{q^2+q'^2}\right] \nonumber\\
&+ \frac{q_i' q_j'}{q'^2}\left[\frac{1}{q^2+ (1+\kappa)q'^2}-\frac{1}{q^2+q'^2}\right] \nonumber \\
&+\frac{q_i q_j' \q \cdot \q'}{q^2 q'^2}\bigg[\frac{1}{q^2+q'^2}+ \frac{1}{(1+\kappa)(q^2+q'^2)} \nonumber\\&- \frac{1}{(1+\kappa)q^2 + q'^2}-\frac{1}{q^2+(1+\kappa)q'^2}\bigg].
\end{align}
Equation~\eqref{eq:uiujElastic} can be used to write a corresponding correlation function for the strain tensor $2 u_{ij}(\q) = -i [q_i u_j(\q) + q_j u_i(\q)]$. Since $H \sim 1/q^2$, the contribution of the tracer to the real-space correlation function of the strain tensor will scale as $1/x^{-2d}$, so that the cold tracer has a long-ranged effect on the fluctuations of the elastic medium. 

As a demonstration, we compute the correlation function of the trace of the strain tensor $u_{ii} = \nabla \cdot \bu$, which describes local compressive deformations of the material. We find
\begin{equation}-\frac{1+\kappa}{T}\langle q_i u_i(\q) q_j' u_j(\q')\rangle =  (2\pi)^d \delta(\q + \q') + \frac{2 \q \cdot \q'}{L^d (q^2 + q'^2)}\end{equation}
The inverse transform of this was computed in Sec.~\ref{sec:coarseGrainedLattice}, Eqs.~\ref{eq:inverseFT} and~\ref{eq:twoPointfield}, leading to the result given in Eq.~\eqref{eq:twoPointElastic}.

\section{Localized inclusions in equilibrium elastic media} \label{appen:equilibriuminclusion}
Here, we revisit the models discussed in Sec.~\ref{sec:fieldtheory} and consider a different perturbation: rather than locally cooling the central lattice node at $\mathbf{r} = 0$, we instead stretch or compress the springs attached to this node through an external potential. If all springs are stretched (compressed) equally, then this is equivalent to pinning the divergence of the displacement field $\nabla \cdot \mathbf{u}|_{\mathbf{r}=0}$ to some positive (negative) value $g$. We let this pinning be harmonic with stiffness $c$ and thus modify the Hamiltonian of Eq.~\eqref{eq:HamLandau} as follows:
\begin{equation} \label{eq:pinnedHamiltonian} 
\mathcal{H} = \int d^{d}\mathbf{r} \left[ \mu (u_{ik})^2 + \frac{1}{2}\lambda (u_{ii})^2 + \frac{1}{2} c \delta(\mathbf{r})(u_{ii} - g)^2\right]\;.
\end{equation}
where we have used $\nabla \cdot \mathbf{u} \equiv u_{ii}$. As this change does not drive the system out of equilibrium, the stationary distribution of the strain field is the Boltzmann distribution $\mathcal{P}[u] \propto e^{-\mathcal{H}[u]/T}$. Notably, the Hamiltonian does not couple the fluctuations of the strain tensor at different locations, so that $u_{ij}(0)$ is independent from any $u_{ij}(\mathbf{r} \neq 0)$. By separating out the traceless part of the strain tensor, we may write a Hamiltonian for $u_{ii}$:
\[\mathcal{H}_\mathrm{tr} = \int d^d\br \left[ \frac{\lambda + 2\mu}{2} (u_{ii})^2 + \frac{c}{2}\delta(\br) (u_{ii}-g)^2\right].\]
The Gaussian correlators may now be computed by inverting the kernel of the quadratic form above, yielding
\begin{align} 
\langle (\nabla \cdot \bu) (r) \rangle =& \frac{c g}{\lambda + 2\mu + c L^d} \delta(\br) \\
\langle (\nabla \cdot \bu)(\br) (\nabla \cdot \bu)(\br')\rangle_c =& \nonumber\frac{T}{\lambda + 2 \mu}\Big[\delta(\br - \br') 
\\ &- \frac{c}{\lambda + 2\mu + c L^d} \delta(\br) \delta(\br')\Big]. \label{eq:nonumber}
\end{align}
Comparing Eq.~\eqref{eq:nonumber} to Eq.~\eqref{eq:twoPointElastic} shows that the long-ranged suppression of fluctuations has been replaced by a purely local effect: Away from $\br = 0$, the correlations are entirely insensitive to the localized potential, showing that local perturbations have purely local effects in equilibrium elastic media. Had there been higher-order gradient terms in Eq.~\eqref{eq:pinnedHamiltonian}, we would have found that the effect of the inclusion decays exponentially with a finite correlation length.  The long-ranged influence of the local inclusion is thus a fundamentally nonequilibrium effect, with analogies in diffusive and active matter systems~\cite{inclusions}.
\bibliography{./bibliography}
\end{document}